\documentclass[sigplan,screen]{acmart}

\usepackage{booktabs}   
\usepackage{subcaption} 

\usepackage{subfiles}
\usepackage{color,amsmath,amssymb,stmaryrd,pifont}
\usepackage{xspace,tikz,wrapfig}
\usepackage{cleveref}
\usepackage{titlecaps}
\usepackage{fancyvrb}
\usepackage{placeins}
\usepackage[multiple]{footmisc}
\usepackage{flushend}

\usepackage[tableposition=top]{caption}

\makeatletter

\setcopyright{acmlicensed}
\acmPrice{15.00}
\acmDOI{10.1145/3192366.3192411}
\acmYear{2018}
\copyrightyear{2018}
\acmISBN{978-1-4503-5698-5/18/06}
\acmConference[PLDI'18]{39th ACM SIGPLAN Conference on Programming Language Design and Implementation}{June 18--22, 2018}{Philadelphia, PA, USA}

\newcommand{\Herbgrind}{Herbgrind\xspace}
\newcommand{\todo}[1]{\relax}

\definecolor{findred}{rgb}{0.82, 0.1, 0.26}
\definecolor{replacegreen}{rgb}{0.53, 0.66, 0.42}

\newcommand{\Mem}{\ensuremath{\mathcal{M}}}
\newcommand{\Err}{\ensuremath{\mathcal{E}}}
\newcommand{\K}[1]{\text{\textbf{#1}}}
\newcommand{\CC}[1]{\text{\textsf{#1}}}

\newcommand{\ZZ}{\ensuremath{\mathbb{Z}}}
\newcommand{\FF}{\ensuremath{\mathbb{F}}}
\newcommand{\RR}{\ensuremath{\mathbb{R}}}
\newcommand{\denote}[1]{\ensuremath{\llbracket #1 \rrbracket}}
\newcommand{\many}[1]{\ensuremath{\overrightarrow{#1}}}
\renewcommand{\>}{\quad\quad}
\newcommand{\GAP}{\vspace{.1in}\\}
\newcommand{\ite}[3]{\ensuremath{\K{if}\:#1\:\K{then}\:#2\:\K{else}\:#3}}
\newcommand{\PC}{\ensuremath{\mathsf{PC}}}
\newcommand{\Addr}{\ensuremath{\mathsf{Addr}}}
\newcommand{\Expr}{\ensuremath{\mathsf{Expr}}}
\newcommand{\addr}[1]{\ensuremath{0{\times}\makeatletter\two@digits{#1}}}
\newcommand{\cons}{\ensuremath{\operatorname{::}}}
\newcommand{\COMMENT}[1]{\ensuremath{//\:\textit{#1}}}
\newcommand{\cmark}{\ding{51}}
\newcommand{\xmark}{\ding{55}}

\newcommand{\CUT}[1]{\relax}

\newcommand{\nDefaultPrecision}{1000\xspace}
\newcommand{\nDepthBound}{5\xspace}

\newcommand{\spot}{spot\xspace}
\newcommand{\spotsnospace}{spots}
\newcommand{\spots}{\spotsnospace\xspace}

\begin{document}
\title{Finding Root Causes of Floating Point Error}

\author{Alex Sanchez-Stern}
\affiliation{
  \institution{University of California San Diego}
  \country{United States of America}
}
\email{alexss@eng.ucsd.edu}

\author{Pavel Panchekha}
\affiliation{
  \institution{University of Washington}
  \country{United States of America}
}
\email{pavpan@cs.washington.edu}

\author{Sorin Lerner}
\affiliation{
  \institution{University of California San Diego}
  \country{United States of America}
}
\email{lerner@cs.ucsd.edu}

\author{Zachary Tatlock}
\affiliation{
  \institution{University of Washington}
  \country{United States of America}
}
\email{ztatlock@cs.washington.edu}


\begin{abstract}

  Floating-point arithmetic plays a central role in science,
  engineering, and finance by enabling developers to approximate
  real arithmetic.
  To address numerical issues in large floating-point applications,
  developers must identify root causes, which is difficult because
  floating-point errors are generally non-local, non-compositional,
  and non-uniform. 

  This paper presents \Herbgrind, a tool
  to help developers identify
  and address root causes in numerical code written in low-level languages like
  C/C++ and Fortran.
  \Herbgrind dynamically tracks dependencies between operations and program outputs to
  avoid false positives and abstracts erroneous computations to
  simplified program fragments whose improvement can reduce output error.
  We perform several case studies applying \Herbgrind to large,
  expert-crafted numerical programs and show that it scales to
  applications spanning hundreds of thousands of lines,
  correctly handling the low-level details of modern floating point hardware
  and mathematical libraries and tracking error across function
  boundaries and through the heap.

\end{abstract}

\ccsdesc[300]{Software and its engineering~Software maintenance tools}

\keywords{floating point, debugging, dynamic analysis}

\maketitle
\section{Introduction}
\label{sec:introduction}

Large floating-point applications play a central role in science,
engineering, and finance by enabling engineers
to approximate real number computations.
Ensuring that these applications provide accurate results (close to the
ideal real number answer) has been a challenge for decades~\cite{kahan-summation,kahan-survey,much-ado-nothing,no-sampling,kahan-future}.
Inaccuracy due to rounding errors has led to market
distortions~\cite{distort-stock, wall-street-distort-stock}, retracted
scientific articles~\cite{num-issues-in-stat, num-replication}, and
incorrect election results~\cite{round-elections}.

Floating-point errors are typically silent:
  even when a grievous error has invalidated a computation, it will
  still produce a result without any indication things have gone awry.
Recent work~\cite{fpdebug,baozhang}
  has developed dynamic analyses that detect
  assembly-level operations with
  large intermediate rounding errors,
  and recent static error analysis tools
  have been developed to verify the
  accuracy of small numerical kernels~\cite{rosa,fptaylor}.

However, after floating-point error has been detected,
  there are no tools to help a developer
  diagnose and debug its \emph{root cause}.
The root cause is the part of a computation whose
  improvement would reduce error in the program's outputs.
Previous work in the area~\cite{fpdebug}
  has called out root cause analysis as a significant open problem
  that needs to be addressed.
This paper addresses this problem
  with a tool that works on large, real,
  expert-crafted floating-point applications
  written in C/C++, and Fortran.

In practical settings, root causes are difficult to identify precisely.
Root causes often involves computations that cross function
boundaries, use the heap, or depend on particular inputs.
As such, even though root causes are part of the computation, they
rarely appear as delineable syntactic entities in the original program
text.
The key challenge, then, is identifying root causes and suitably
abstracting them to a form that enables numerical analysis
and facilitates improving the accuracy of the program.
\CUT{To the best of our knowledge, no approach exists to help developers
identify and address root causes in numerical code written in
low-level C/C++ and Fortran programs spanning hundreds of thousands of
lines.}

From a developer's perspective, identifying and debugging numerical
issues is difficult for several reasons.
First, floating-point error is non-compositional: large intermediate
errors may not impact program outputs and small intermediate errors
may blow up in a single operation (due to, e.g.,~cancellation or overflow).
Second, floating-point errors are often non-local: the source of an error
can be far from where it is observed and may involve values that cross
function boundaries and flow through heap-allocated data structures.
Third, floating-point errors are non-uniform: a program may have high
error for certain inputs despite low or non-existent error on
other inputs.

This paper presents \Herbgrind, a dynamic binary analysis that
identifies \textit{candidate root causes} for numerical error in large
floating-point applications.
%
%
\CUT{\Herbgrind uses a variety of novel techniques
to overcome the non-compositional, non-local, non-uniform nature
of floating-point error and pinpoint the root causes of error
in large programs.}
First, to address the non-compositionality of error,
  \Herbgrind records operations with intermediate error
  and tracks the \emph{influence} of these operations
  on program outputs and control flow.
Second, to address the non-locality of error,
  \Herbgrind provides symbolic expressions
  to describe erroneous computation,
  abstracting the sequence of operations that
  cause error to program fragments which
  facilitate numerical analysis.
Finally, to address the non-uniformity of error,
  \Herbgrind characterizes the inputs to erroneous computations
  observed during analysis, including the full range as well as
  the subset that caused significant error.

We demonstrate \Herbgrind's effectiveness by identifying the root
causes of error in three expert-written numerical applications and
benchmarks.
We find that \Herbgrind handles the tricky floating-point
manipulations found in expert-written numerical code, and can identify
real sources of floating-point error missed by experts when writing
important software.
To further characterize the impact of \Herbgrind's key components,
we carried out a set of smaller experiments with the FPBench
floating-point benchmark suite~\cite{fpbench}, and find that each of
\Herbgrind's components is crucial to its accuracy and performance.

To the best of our knowledge, \Herbgrind provides the first approach to
identifying and summarizing root causes of error in large numerical programs.
\Herbgrind is implemented in
  the Valgrind binary instrumentation framework,
  and achieves acceptable performance
  using several key optimizations.\footnote{The implementation of
  \Herbgrind is publicly available at \\ \url{https://github.com/uwplse/herbgrind}}
Building \Herbgrind required developing the following contributions:
\begin{itemize}

\item An analysis that identifies candidate root causes by tracking
  dependencies between sources of error and program outputs,
  abstracting the responsible computations to an improvable program
  fragment, and characterizing the inputs to these computations
  (\Cref{sec:spec}).

\item An implementation of this analysis that supports numerical code
  written in low-level languages like C/C++ and Fortran and handles the
    complexities of modern floating point hardware and
    libraries (\Cref{sec:impl}).

\item Key design decisions and optimizations required for this
  implementation to achieve acceptable performance when scaling up to
    applications spanning hundreds of thousands of lines
    (\Cref{sec:opt}).

\item An evaluation of \Herbgrind including bugs found ``in the wild''
  and measurements of the impact of its various subsystems (\Cref{sec:cases} and
  \Cref{sec:eval}).
\end{itemize}

\section{Background}
\label{sec:background}

A floating-point number represents a real number of
the form \( \pm (1 + m) 2^e\), where $m$ is a fixed-point value
between 0 and 1 and $e$ is a signed integer; several other values,
including two zeros, two infinities, not-a-number error values, and
subnormal values, can also be represented. In double-precision
floating point, $m$ is a 52-bit value, and $e$ is an 11-bit value,
which together with a sign bit makes 64 bits. Simple operations on
floating-point numbers, such as addition and multiplication, are
supported in hardware on most computers.

\subsection{Floating-Point Challenges}
\label{ssec:challenges}

\paragraph{Non-compositional error}
Individual floating-point instructions are always evaluated
  as accurately as possible, but since not all real numbers
  are represented by a floating-point value,
  some error is necessarily produced.
Thus, the floating-point sum of $x$ and $y$
  corresponds to the real number $x + y + (x + y)\epsilon$,
  where $\epsilon$ is some small value induced by rounding error.%
\footnote{For cases involving subnormal numbers,
  the actual error formula is more complex;
  these details are elided here.}
Floating-point operations implemented in libraries
  also tend to bound error to a few units in the last place
  (ulps) for each operation,
  but are generally not guaranteed to be the closest result.
However error grows when multiple operations are performed.
For example, consider the expression $(x + 1) - x = 1$.
The addition introduces error $\epsilon_1$
  and produces $x + 1 + (x + 1)\epsilon_1$.
The subtraction then introduces $\epsilon_2$ and produces
\begin{align*}
  1 + (x + 1) \epsilon_1 + \epsilon_2 + (x + 1) \epsilon_1 \epsilon_2.
\end{align*}
Since $x$ can be arbitrarily large, the $(x + 1) \epsilon_1$ term can be large;
  in fact, for values of $x$ on the order of $10^{16}$, the expression $(x + 1) - x$
  evaluates to 0, not 1.
\CUT{The composition of relatively accurate addition and subtraction
  operations can produce very inaccurate results~\cite{kahan-java-hurts,pitfalls}.
Analogs of this situation occur often in the real world~\cite{pitfalls},
  where they can be difficult to detect.}
The influence of intermediate errors on program output can be subtle;
  not only can accurate intermediate operations compute an inaccurate result,
  but intermediate error can cancel to produce an accurate result.
Experts often orchestrate such cancellation,
  which poses a challenge to dynamic analysis tools
  trying to minimize false positives.

\CUT{Past work~\cite{fpdebug} investigated
a floating-point error in the SPEC benchmark CalculiX, and
concluded that ``the influence may be negligible for this
simulation'' because the erroneous operation did not affect the
program's result.}

\paragraph{Non-local error}
Floating-point error can also be non-local; the cause of a floating
point error can span functions and thread through data structures.
Consider the snippet:
\begin{verbatim}
  double foo(struct Point a, struct Point b) {
    return ((a.x + a.y) - (b.x + b.y)) * a.x;
  }
  double bar(double x, double y, double z) {
    return foo(mkPoint(x, y), mkPoint(x, z));
  }
\end{verbatim}
The \CC{foo} and \CC{bar} functions individually appear accurate.
However, \CC{bar}'s use of \CC{foo} causes inaccuracy.
For example for inputs \CC{x=1e16}, \CC{y=1}, \CC{z=0},
  the correct output of \CC{bar} is \CC{1e16},
  yet \CC{bar} instead computes \CC{0}.
However, this combination of \CC{foo} and \CC{bar}
  can be computed more accurately,
  with the expression $(y - z) \cdot x$.
Note that in this example, understanding the error
  requires both reasoning across function boundaries
  and through data structures.


\paragraph{Non-uniform error}
For a given computation, different inputs can
  cause vastly different amounts of floating-point error.
Effectively debugging a numerical issue
  requires characterizing the inputs
  which lead to the root cause impacting output accuracy.
For example, consider the snippet:
\begin{verbatim}
  double baz(double x){
    double z =  1 / (x - 113);
    return (z + M_PI) - z;
  }
\end{verbatim}
When debugging \CC{baz}, it's important to know what inputs \CC{baz}
is called on in the context of the larger program.
For most inputs, \CC{baz} is accurate; if \CC{baz} is only called on
inputs far from $113$, then a programmer need not consider it
problematic.
However, for values of \CC{x} near $113$, \CC{baz} suffers significant
rounding error, because z becomes very large, and then most of the
bits of $\pi$ are lost to catastrophic cancellation.
To diagnose the root cause of error in a program containing \CC{baz},
programmers need to know whether \CC{baz} is called on inputs near
$113$; if not, they may waste time investigating \CC{baz}'s behavior
on inputs near $113$ when those inputs are never seen in practice.
In this example, understanding the error requires reasoning about the
inputs on which the fragment of code will be executed.

\subsection{Existing Debugging Tools}

\begin{table*}
  \caption{Comparison of floating-point error detection tools. Note
    that all tools are run on distinct benchmark suites.}
  \begin{tabular}{l  c  c  c  c }
    Feature & FpDebug & BZ & Verrou & \Herbgrind \\
    \hline
    \multicolumn{5}{l}{\textbf{Error Detection}}\\
    \hspace{0.5cm}Dynamic & \cmark & \cmark & \cmark & \cmark \\
    \hspace{0.5cm}Detects Error& \cmark & \cmark & \cmark & \cmark \\
    \hspace{0.5cm}Shadow Reals& \cmark & \xmark & \xmark & \cmark \\
    \hspace{0.5cm}Local Error&\xmark & \xmark & \xmark & \cmark \\
    \hspace{0.5cm}Library Abstraction& \xmark & \xmark & \xmark & \cmark \\
    \multicolumn{5}{l}{\textbf{Root Cause Analysis}}\\
    \hspace{0.5cm}Output-Sensitive Error Report& \xmark & \xmark & \xmark & \cmark \\
    \hspace{0.5cm}Detect Control Divergence & \xmark & \cmark & \xmark & \cmark \\
    \hspace{0.5cm}Localization & Opcode Address & None & None & Abstracted Code Fragment \\
    \hspace{0.5cm}Characterize Inputs & \xmark & \xmark & \xmark & \cmark \\
    \multicolumn{5}{l}{\textbf{Other}}\\
    \hspace{0.5cm}Automatically Re-run in High Precision & \xmark & \cmark & \xmark & \xmark \\
    \hspace{0.5cm}Overhead* & 395x & 7.91x & 7x & 574x \\
  \end{tabular}

  \label{fig:feature-table}
\end{table*}

\Cref{fig:feature-table} compares the tools most closely related to \Herbgrind.

\paragraph{Error detection}
All of the tools compared are dynamic analyses which attempt to detect error.
Like \Herbgrind, \newline FpDebug\cite{fpdebug} uses high-precision shadow values to
track the error of floating-point computations; Verrou\cite{verrou} and
BZ\cite{baozhang} use heuristic methods to detect possible instances of error.

While individual hardware floating-point operations are accurate, more complex
operations, like trigonometric functions, are generally implemented in
low-level math libraries which make use of hundreds of floating-point
instructions and bit manipulations for each operation, and are painstakingly
crafted by experts~\cite{fdlibm, openlibm}. Even higher level operations, like
those on matrices and vectors, are implemented in thousands of lower-level
operations, often building on both hardware floating-point instructions and
lower-level math libraries~\cite{blas}.
Previous tools report error locations as individual opcode addresses, which may
be deep within the internals of the implementation of a sophisticated
operation.
This is unfortunate, as most users are unwilling and/or unable to modify a
\texttt{libm} or BLAS implementation; instead such operations should be treated
as atomic black boxes so that users can focus on ensuring such operations are
accurately used at a higher level of abstraction.
In contrast, \Herbgrind supports abstracting over such
library calls which enables more useful error location reporting and permits
meaningful shadowing in high-precision.

\paragraph{Root cause analysis}
There are two ways in which floating-point error can affect observable
program behavior: either by flowing directly to an output, or changing
control flow.
BZ~\cite{baozhang} can detect changes in control flow due to error; but
cannot reason about how error affects program outputs.
Verrou and FpDebug have no support for detecting observable program
behavior affected by error.
When an error is detected, FpDebug reports the binary address where
the error was found, while Verrou and BZ only report that something
has gone wrong, not where it has gone wrong.
In contrast, \Herbgrind characterizes both the full range
and error-inducing inputs to the problematic computations
observed during analysis.





\section{Overview}
\label{sec:overview}

\begin{figure}
  \includegraphics[width=.49\linewidth]{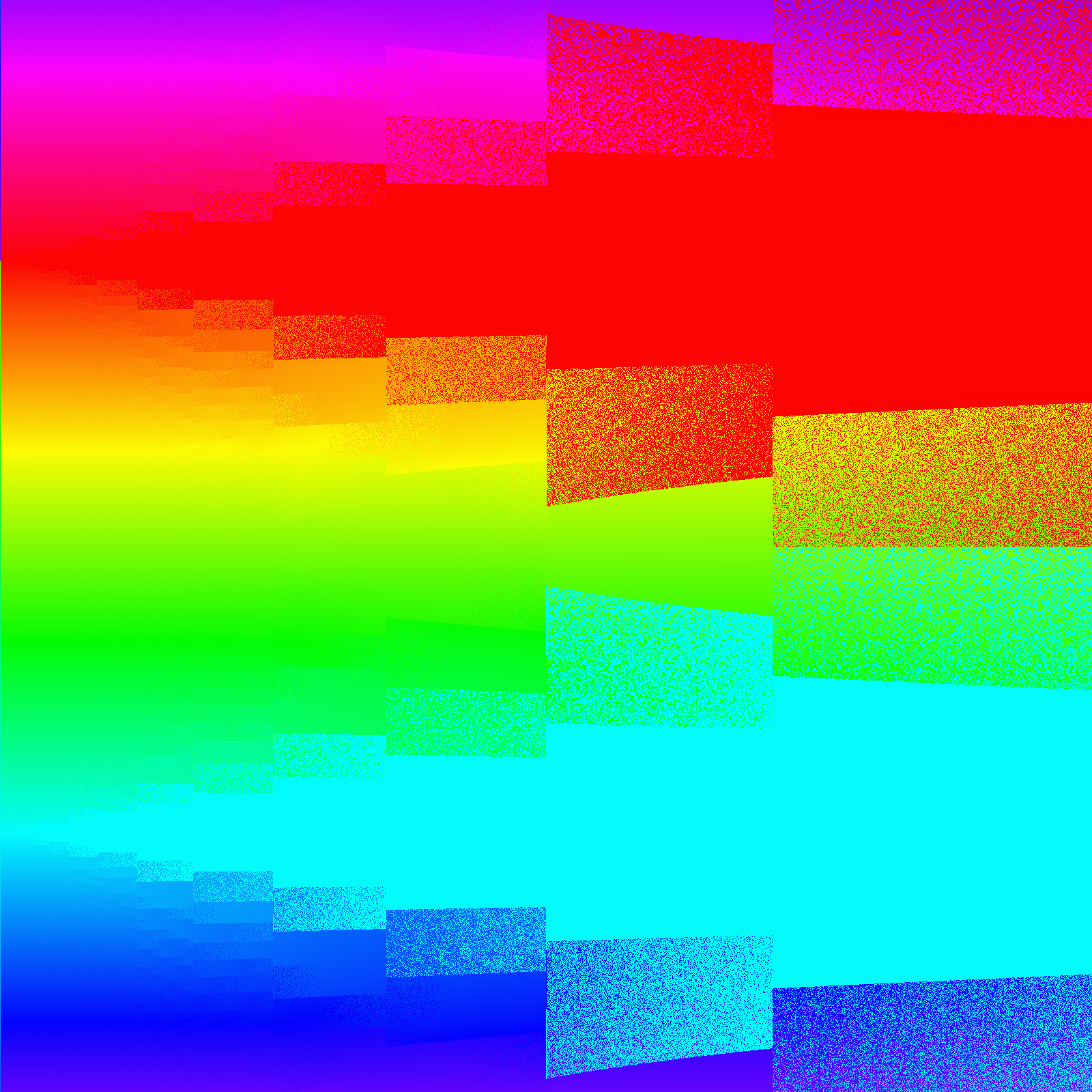}%
  \hfill%
  \includegraphics[width=.49\linewidth]{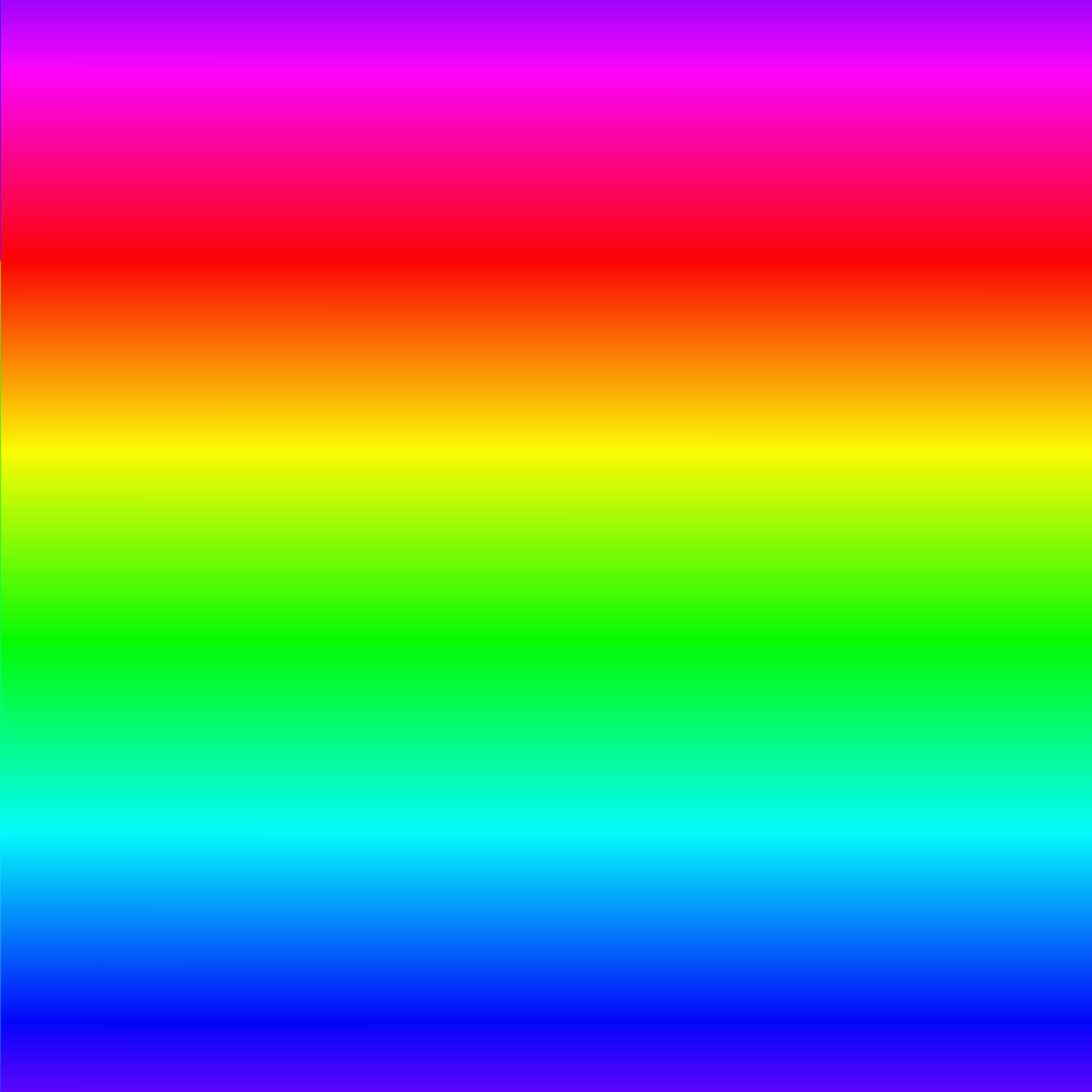}
  \caption{Complex plotter output
    before (left) and after (right) diagnosing and fixing
    a floating-point error.}
  \label{fig:ovw-output}
\end{figure}



We illustrate \Herbgrind by detailing its use
  on a simple complex function plotter.
Given function $f : \mathbb{C} \to \mathbb{C}$,
  region $R = [x_0, x_1] \times [y_0, y_1]$ in the complex plane,
  and a resolution,
  the plotter tiles $R$ with a grid of pixels,
  and colors each pixel based on $\arg(f(x + i y))$
  at the center $(x, y)$ of each pixel.%
\footnote{\texttt{arg} is also known as \texttt{atan2}}%
\CUT{The program spends the majority of its time evaluating $f$,
  and uses standard data structures for complex numbers,
  libraries for the manipulation of bitmap images,
  and bitwise manipulation of colors.}
Since the output of the program is an image,
  minor errors in the evaluation of $f$
  can usually be ignored.
However, floating-point error can compound in unexpected ways.
Consider the function\footnote{$\Re(z)$ indicates the real (non-imaginary) part of $z$}
\[
f(z) = 1 / \left( \sqrt{\Re(z)} - \sqrt{\Re(z) + i \exp(-20 z)} \right).
\]
To evaluate this function,
  the plotter must provide codes
  for evaluating the square root of a complex number.
The standard formula is
\[
\sqrt{x + i y} = \left( \sqrt{\sqrt{x^2 + y^2} + x} + i \sqrt{\sqrt{x^2 + y^2} - x} \right) / \sqrt{2}, \\
\]
  where the square roots in the definitions
  are all square roots of real numbers
  (typically provided by standard math libraries).

%
Implementing $f$ using this formula
  and plotting the region $R = [0, \frac14] \times [-3, 3]$
  results in the left image of \Cref{fig:ovw-output}.
The speckling is not an accurate representation of $f$;
  in fact, $f$ is continuous in both $x$ and $y$ throughout $R$.




\Herbgrind uses three key components
  (detailed in \Cref{sec:spec})
  to identify the root cause of error
  in the plotting program:
(1) a shadow taint analysis, which tracks the ways
  that erroneous operations influence important program locations
  called \textit{\spots};
(2) a shadow symbolic execution, which builds
  expressions representing the computations that
  produced each value; and
(3) an input characterization system, which
  maintains information about the set of inputs to each computation.

\Herbgrind detects that the plotter computes
  wrong pixel values due to significant error from a
  subtraction with high local error:

\vspace{0.1in}
{\centering
\footnotesize
\begin{BVerbatim}
  Compare @ main.cpp:24 in run(int, int)
  231878 incorrect values of 477000
  Influenced by erroneous expressions:
    (FPCore (x y)
      :pre (and (<= -2.061152e-9 x 2.497500e-1)
                (<= -2.619433e-9 y 2.645912e-9))
          (- (sqrt (+ (* x x) (* y y))) x))
   Example problematic input: (2.061152e-9, -2.480955e-12)
\end{BVerbatim}
}
\vspace{0.1in}

\noindent
This report shows that at line 24 of the \texttt{main.cpp} source file,
inaccuracy is caused by the expression:
\[
  \sqrt{x^2 + y^2} - x.
\]
%

Running Herbie~\cite{herbie} on the above expression
produces this more accurate version.
\[
\sqrt{x^2 + y^2} - x \rightsquigarrow
\begin{cases}
  \sqrt{x^2 + y^2} - x & \K{if}\:x \le 0 \\[6pt]
  y^2 / \left(\sqrt{x^2 + y^2} + x\right) & \K{if}\:x > 0 \\[6pt]
\end{cases}.
\]
Substituting this expression back into
the original complex square root definition
(and simplifying) yields
{\small
\[
\sqrt{x+iy} =
\frac1{\sqrt2}
\begin{cases}
  |y| / \sqrt{\sqrt{x^2 + y^2} - x} + i \sqrt{\sqrt{x^2 + y^2} - x} & \K{if}\:x \le 0 \\[6pt]
  \sqrt{\sqrt{x^2 + y^2} + x} + i  |y| / \sqrt{\sqrt{x^2 + y^2} + x} & \K{if}\:x > 0 \\[6pt]
\end{cases}
\]
}
for the complex square root.
Replacing the complex square root computation in the plotter
  with this alternative implementation
  fixes the inaccurate computation,
  as confirmed by running \Herbgrind on the repaired program%
  \footnote{Automating the process of inserting improved code back
    into the binary is left to future work, but \Herbgrind can provide
    source locations for each node in the extracted expression.}.
The fixed code produces the right graph in \Cref{fig:ovw-output}.

While error in the complex plotter is primarily in a single compilation unit, real
world numerical software often has many numerical components which interact,
and root causes often cannot be isolated to one component. We evaluate
\Herbgrind on several large numerical systems which exhibit these issues in
\Cref{sec:cases}.


\section{Analysis}
\label{sec:spec}

The implementation of \Herbgrind's analysis
  requires handling complex machine details
  including different value precisions, multiple distinct storage types, and
  bit-level operations, which we detail in \Cref{sec:impl}.
To first clarify the core concepts,
  this section describes \Herbgrind's analysis
  in terms of an abstract float machine.
\CUT{Since root causes are program fragments
  which influence program outputs with significant error,
  \Herbgrind's analysis computes the error of program outputs,
  finds the erroneous computations that influence those outputs,
  builds expressions to describe those computations,
  and characterizes the inputs to those expressions.}
\Herbgrind's analysis consists of three components:
  a \spots-and-influences system
  to determine which operations influence which program outputs
  (\Cref{ssec:marks-and-influences}),
  a symbolic expression system
  to track computations across function and heap data structure boundaries
  (\Cref{ssec:symbolic-shadows}),
  and an input characteristics system
  to determine on which inputs the computation is erroneous or accurate
  (\Cref{ssec:input-characteristics}).
\CUT{Each component
  shadows floating-point values in memory and
  updates its shadow memory and operation information
  for each floating-point operation
  executed by the program (\Cref{fig:components}).}

\subsection{Abstract Machine Semantics}
\label{ssec:semantics}

The abstract machine has floating-point values and operations
  as well as memory and control flow~(\Cref{fig:machine}).
A machine contains mutable memory
  $\Mem : \ZZ \to (\FF \;|\; \ZZ)$
  which stores floating-point values or integers
  and a program counter $\CC{pc} : \ZZ$
  that indexes into the list of program statements.
\CUT{We define operations and memory as typed:
  a memory location stores either floating-point values or integers,
  but cannot store both at different times,
  and each operation has a fixed set of floating-point and integer inputs.}
Statements include: computations, control operations, and outputs.
\CUT{Computations read values from memory locations,
  evaluate a function on those inputs,
  and write the results back to memory;
  control operations read values from memory locations,
  evaluate a predicate on those values,
  and jump to a new program counter;
  and output operations read a value from memory
  and use it to perform I/O.}
A program is run by initializing the memory and program counter,
  and then running statements until the program counter becomes negative.

\begin{figure*}
  {
    \small
  \begin{subfigure}[t]{.48\linewidth}
  \[\begin{array}{l}
    \Addr = \PC = \ZZ \\
    \Mem[a : \Addr] : \FF \mid \ZZ \GAP
    \CC{pc} : \PC \\

    \CC{prog}[n : \PC] : \Addr \gets f(\many{\Addr})\\
    \phantom{\CC{prog}[n : \PC]}
    \mid\K{if}\:P(\many{\Addr})\:\K{goto}\:\PC\\
    \phantom{\CC{prog}[n : \PC]}
    \mid\K{out}\:\Addr \GAP

    \CC{run}(\CC{prog}) =\\
    \>\Mem[-] = 0, \CC{pc} = 0 \\
    \>\K{while}\:\CC{pc} \ge 0\\
    \>\>\Mem, \CC{pc} = \denote{\CC{prog}[\CC{pc}]}(\Mem, \CC{pc})
  \end{array}\]
  \end{subfigure}
  \hfill
  \begin{subfigure}[t]{.48\linewidth}
  \[\begin{array}{l}
    \denote{y \gets f(\many{x})}(\Mem, \CC{pc}) =\\
    \>r = \denote{f}(\many{\Mem[x]})\\
    \>\Mem[y \mapsto r], \CC{pc} + 1 \GAP
    \denote{\K{if}\:P(\many{x})\:\K{goto}\:n}(\Mem, \CC{pc}) =\\
    \>\CC{pc}' = \ite{\denote{P}(\many{\Mem[x]})}{n}{\CC{pc} + 1}\\
    \>\Mem, \CC{pc}' \GAP
    \denote{\K{out}\:x}(\Mem, \CC{pc}) =\\
    \>\CC{print}\:\Mem[x] \\
    \>\Mem, \CC{pc} + 1
  \end{array}\]
  \end{subfigure}
  }
  \caption{
    The abstract machine semantics for low-level floating-point computation.
    A machine contains memory and a program counter which
    indexes into a list of statements.
    Statements either compute values and store the result in memory,
      perform a conditional jump,
      or output a value.}
  \label{fig:machine}
\end{figure*}

\Herbgrind's analysis describes candidate root causes
  for programs on this abstract machine
  (\Cref{fig:analysis} and \Cref{fig:components})
  by updating analysis information
  for each instruction in the program.
Each computation instruction with floating-point output
  is associated with an operation entry,
  which describes the computation that led up to that operation,
  and a summary of the values that computation takes.
All other instructions have a \spot entry,
  which lists the error at that location
  and the erroneous computations that influence the location
  (\Cref{ssec:marks-and-influences}).

\begin{figure*}[t]
  {
    \small
  \begin{subfigure}[t]{.45\linewidth}
  \[\begin{array}{l}
    \Expr = \RR \mid f(\many{\Expr}) \GAP

    \Mem_\RR[a : \Addr] : \RR \\
    \Mem_I[a : \Addr] : \CC{Set}\:\PC \\
    \Mem_E[a : \Addr] : \Expr \GAP

    \CC{ops}[n : \PC] : \\
    \hspace{.5cm}\CC{Set}\:\left(\CC{Expr}\:\times\:\CC{List}\left(\CC{Set}\:\FF\right)\:\times\:\left(\CC{Position}\to\CC{Set}\:\FF\right)\right) \\
    \CC{\spots}[n : \PC] : (\CC{Set}\:\RR) \times (\CC{Set}\:\PC)
  \end{array}\]
  \end{subfigure}
  \hfill
  \begin{subfigure}[t]{.53\linewidth}
  \[\begin{array}{l}
    \CC{analyze}(\CC{prog}) =\\
    \>\Mem[-] = \Mem_\RR[-] = \Mem_E[-] = 0,\Mem_I[-] = \emptyset\\
    \>\CC{pc} = 0, \CC{\spots}[-] = (\emptyset, \emptyset), \CC{ops}[-] = \emptyset\\
    \>\K{while}\:\CC{pc} \ge 0\\
    \>\>\Mem', \CC{pc}' = \denote{\CC{prog}[\CC{pc}]}(\Mem, \CC{pc}) \\
    \>\>\Mem_\RR' =
      \denote{\CC{prog}[\CC{pc}]}_\RR(\Mem, \Mem_\RR)\\
    \>\>\Mem_I' =
      \denote{\CC{prog}[\CC{pc}]}_I(\Mem_\RR, \Mem_I, \CC{pc}) \\
    \>\>\Mem_E' =
      \denote{\CC{prog}[\CC{pc}]}_E(\Mem_\RR', \Mem_E) \\
    \>\>\CC{record}(\CC{prog}, \CC{pc}, \CC{ops}, \CC{\spots}, \Mem', \Mem_\RR, \Mem_I', \Mem_E')\\
    \>\>\CC{pc} = \CC{pc}', \Mem = \Mem', \Mem_\RR = \Mem_\RR', \Mem_I =
    \Mem_I', \Mem_E = \Mem_E' \\
    \>\K{return}\:\CC{\spots}, \CC{ops}
  \end{array}\]
  \end{subfigure}
  }
  \caption{The \Herbgrind analysis for finding root causes for floating-point error.
    \Herbgrind maintains shadow memories
    for real values ($\Mem_\RR$), influences ($\Mem_I$), and concrete expressions ($\Mem_E$).
    Additionally, \Herbgrind tracks
    concrete expressions and input sets (both total and problematic)
    for operations in \CC{ops}
    and error and influences for \spots in \CC{\spots}.
    Note that \Herbgrind follows floating-point control flow branches
    during analysis; cases when it diverges from the control flow interpreted
    under reals are reported as errors.
  }
  \label{fig:analysis}
\end{figure*}

\begin{figure*}
  {
    \small
  \begin{subfigure}[t]{.48\linewidth}
  \[\begin{array}{l}
    \COMMENT{Reals} \\
    \denote{y \gets f(\many{x})}_\RR(\Mem, \Mem_\RR)\:\K{when}\:\Mem_\RR[y] \in \RR=\\
    \>\many{v} = \ite{\many{\Mem_\RR[x]} \in \RR}{\many{\Mem_\RR[x]}}{\many{\Mem[x]}}\\
    \>\Mem_\RR[y \mapsto \denote{f}_\RR(\many{v})] \GAP\GAP

    \COMMENT{Influences} \\
    \denote{y \gets f(\many{x})}_I(\Mem_\RR, \Mem_I, \CC{pc})\:\K{when}\:\Mem_\RR[y] \in \RR=\\
    \>s = \bigcup \many{\Mem_I[x]}\\
    \>\K{if}\:\CC{local-error}(f, \many{\Mem_\RR[x]}) > T_\ell\:\K{then}\\
    \>\>s = \CC{pc} \cons s\\
    \>\Mem_I[y \mapsto s] \GAP

   \CC{local-error}(f, \many{v}) =\\
   \>r_\RR = \FF(\denote{f}_\RR(\many{v})) \\
   \>r_\FF = \denote{f}_\FF(\many{\FF(v)}) \\
   \>\Err(r_\RR, r_\FF)\GAP\GAP

   \COMMENT{Expressions} \\
    \denote{y \gets f(\many{x})}_E(\Mem_\RR, \Mem_E)\:\K{when}\:\Mem_\RR[y] \in \RR=\\
    \>e = \ite{\many{\Mem_\RR[x]} \in \RR}{f(\many{\Mem_E[x]})}{\Mem_\RR[y]} \\
    \>\Mem_E[y \mapsto e]

  \GAP
  \>\CC{update-problematic-inputs}(e, \hat{c})\\
  \>\>\CC{nodes},\:\CC{positions} = \\
  \>\>\>\CC{get-all-descendant-nodes}(e)\\
  \>\>\hat{c}' = \CC{make-table}()\\
  \>\>\K{for}\:\CC{node},\:\CC{position}\:\K{in}\:\CC{zip}(\CC{nodes},\:\CC{positions}):\\
  \>\>\>\hat{c}'[\CC{position}] = \hat{c}[\CC{position}] + \left\{\CC{node.value}\right\}\\
  \end{array}\]
  \end{subfigure}
  \hfill
  \begin{subfigure}[t]{.48\linewidth}
  \[\begin{array}{l}
  \CC{record}(\CC{prog}, \CC{pc}, \CC{ops}, \CC{\spots}, \Mem, \Mem_\RR, \Mem_I, \Mem_E) = \\
  \> e, \many{c},\hat{c} = \CC{ops}[\CC{pc}] \\
  \> \varepsilon, i = \CC{\spots}[\CC{pc}] \\
  \> \K{match}\:\CC{prog}[\CC{pc}]\:\K{with}\\

  \> \mid (y \gets f(\many{x}))\:\K{when}\:\Mem[y] \in \FF \implies \\
  \>\> e' = \Mem_E[y] \cons e \\
  \>\> \many{c}' = \CC{update-total-inputs}(\many{x},\many{c})\\
  \>\> \K{if}\:\CC{local-error}(f, \many{\Mem_\RR[x]}) > T_\ell\:\K{then}\\
  \>\>\> \hat{c}' = \CC{update-problematic-inputs}(\Mem_E[y], \hat{c})\\
  \>\> \CC{ops}[\CC{pc}] = (e',\many{c}',\hat{c}')\\

  \> \mid (y \gets f(\many{x}))\:\K{when}\:\Mem[y] \in \ZZ \implies \\
  \>\> \K{if}\:\denote{f}_\RR(\many{\Mem_\RR[x]}) = \Mem[y]\:\K{then} \\
  \>\>\> \CC{\spots}[\CC{pc}] = (1 \cons \varepsilon, i \bigcup \many{\Mem_I[x]}) \\
  \>\> \K{else} \\
  \>\>\> \CC{\spots}[\CC{pc}] = (0 \cons \varepsilon, i) \\

  \> \mid (\K{if}\:P(\many{x})\:\K{goto}\:y) \implies \\
  \>\> \K{if}\:\denote{P}_\RR(\many{\Mem_\RR[x]}) = \denote{P}(\many{\Mem[x]})\:\K{then} \\
  \>\>\> \CC{\spots}[\CC{pc}] = (1 \cons \varepsilon, i \bigcup \many{\Mem_I[x]}) \\
  \>\> \K{else} \\
  \>\>\> \CC{\spots}[\CC{pc}] = (0 \cons \varepsilon, i) \\

  \> \mid (\K{out}\:x) \implies \\
  \>\> r = \Err(\Mem[x], \Mem_\RR[x]) \\
  \>\> \K{if}\: r > T_m\:\K{then} \\
  \>\>\> \CC{\spots}[\CC{pc}] = (r \cons \varepsilon, i \cup \Mem_I[x]) \\
  \>\> \K{else} \\
  \>\>\> \CC{\spots}[\CC{pc}] = (r \cons \varepsilon, i)

  \GAP

  \>\CC{update-total-inputs}(\many{v},\many{c}) = \\
  \>\>\many{c}' = []\\
  \>\>\K{for}\:v,\:c\:\K{in}\:\CC{zip}(\many{v},\many{c}):\\
  \>\>\>\many{c}' = (c + \left\{v\right\}) :: \many{c}'\\
  \>\>\many{c}' = \CC{reverse}(\many{c}')\\
  \end{array}\]
  \end{subfigure}
  }
  \caption{
    On the left, the real-number execution, influence propagation,
    and concrete expressions building in \Herbgrind;
    shadow executions not shown are no-ops.
    On the right, how \Herbgrind updates
    the operation and \spot information on every statement.
    Below are helper functions.
  }
  \label{fig:components}
\end{figure*}

\paragraph{Shadow Reals}
Floating-point errors are generally silent:
  even when error invalidates a computation,
  that computation still produces a floating-point value
  without any indication that the value is erroneous.
\Herbgrind follows prior work
  in detecting the floating-point errors in a program
  by computing a real number shadow
  for every floating-point value in ordinary program memory.%
\footnote{While the abstract analysis is defined in terms of computing over reals,
  the implementation must settle merely for high precision
  (e.g., 1000-bit mantissa) approximations (\Cref{ssec:values}).}
For each such statement,
  \Herbgrind executes the statement's
  operation in the reals on real-number shadow inputs,
  and stores the resulting real number
  to the real-number shadow memory.
\CUT{Divergences between the floating-point and real-number executions
  are handled by the \spots-and-influences system (\Cref{ssec:marks-and-influences}).}
\Herbgrind's handling of mathematical libraries and similar details
  is discussed in \Cref{sec:impl}.

\subsection{\titlecap{\spotsnospace}\xspace and Influence Shadows}
\label{ssec:marks-and-influences}

\Herbgrind uses the real-number execution
  to measure the floating-point error
  at program outputs, conditional branches,
  and conversions from floating-point values to integers;
  these three types of program locations
  are collectively called \textit{\spots}.
Since error is non-compositional,
  the root cause of error at a \spot can be far from the \spot itself.
To overcome this, \Herbgrind identifies candidate root causes
  and tracks their influence on \spots using a taint analysis:
  every floating-point value has a ``taint'' set of influencing instructions
  which is propagated by computation.

\CUT{Each \spot can see erroneous results:
  an output instruction can output the wrong floating-point value,
  a control instruction can take the wrong branch,%
\footnote{If a control instruction has different results,
  \Herbgrind will follow the floating-point execution by default,
  but this behavior can be controlled by a flag.}
  and any computation of an integer can compute the wrong one.
In the first case,
  \Herbgrind measures the error as the difference in output values;
  in the other cases, error is boolean: correct or incorrect.
Since an instruction can execute multiple times,
  \Herbgrind tracks error for each execution of each \spot.}

\Herbgrind uses \textit{local error} to determine
  which floating-point operations cause error
  (\Cref{fig:components}).
Local error~\cite{herbie}
  measures the error an operation's output would have
  even if its inputs were accurately computed,
  and then rounded to native floats.\CUT{:
\[
\CC{local-error}(f, \many{v}) = \Err(\FF(\denote{f}_\RR(\many{v})), \denote{f}_\FF(\many{\FF(v)}))
\]
Local error allows identifying individual operations
  that can cause errors, even for operations whose inputs are already erroneous.}
Using local error to assess operations avoids blaming innocent operations
  for erroneous operands.
Any operation whose local error passes a threshold $T_\ell$
  is treated as a candidate root cause.%
\footnote{\Herbgrind only reports those sources of error where error
  flows into spots, so that users are only shown erroneous code that
  affects program results.}

In the case of the complex plotter from \Cref{sec:overview},
  the subtraction at the top of the reported expression
  was determined to have high local error,
  resulting in it being tracked.

\subsection{Symbolic Expressions}
\label{ssec:symbolic-shadows}

Identifying the operation that introduced error,
  as the influences system does, is not sufficient to understand the error,
  because floating-point error is non-local:
  to understand why an operation is erroneous
  requires understanding how its inputs were computed.
In many cases those inputs are separated from the erroneous operation
  by function boundaries and heap data structure manipulations.
\Herbgrind analyzes through those boundaries
  by providing \emph{symbolic expressions} for the erroneous operation and its inputs.

Symbolic expressions represent an abstract computation
  that contains a candidate root cause.
Each symbolic expression contains only floating-point operations:
  it therefore abstracts away function boundaries and heap data structures.
A symbolic expression must be general enough
  to encompass any encountered instance of a computation,
  while still being specific enough
  to be as helpful as possible to the user.

\Herbgrind constructs symbolic expressions
  by first recording a \textit{concrete} expression
  for every floating-point value
  and then using \textit{anti-unification}
  to combine these concrete expressions into symbolic expressions.
Each concrete expression
  tracks the floating-point operations that were used to build a
  particular value.
Concrete expression are copied when a value is
  passed to a function, entered in a heap data structure, or stored and
  retrieved later, but these operations themselves are not recorded.
A single concrete expression (and the resulting symbolic expression),
  might encompass a subtraction from one function, a series of
  multiplications from another, and an exponentiation which occurs after
  the value is stored in a hash table and later retrieved.

From the concrete expression for every value,
  \Herbgrind computes symbolic expressions using a variant
  of the classic \textit{anti-unification} algorithm~\cite{antiunification}
  for computing the most-specific generalization of two trees.
Symbolic expressions are much like concrete expressions,
  but include variables which can stand in for any subtree;
  variables which stand in for equivalent subtrees are the same.
To produce expressions which are more useful for program improvement,
  \Herbgrind uses a modified version of anti-unification.
These modifications are described in an extended tech report.
Reporting detailed symbolic expressions is essential
  for diagnosing the root causes of error;
  in the case of the plotter,
  the full extracted expression was essential for producing
  an improved complex square root definition.
\CUT{In both cases, the difference hinges
  on which concrete expressions are considered equivalent.
Since a symbolic expression represents an abstract computation
  it may contain variables for inputs to the computation;
  anti-unification generates variables
  when combining two concrete expressions with different structure.
However,
  variables not only indicate varying concrete expressions
  but also, when a variable repeats, equivalent ones.
Standard anti-unification requires concrete expressions
  to compute the same value in the same way to be equivalent.
\Herbgrind makes two modifications to this requirement.}

\subsection{Input Characteristics}
\label{ssec:input-characteristics}

Because floating-point error is non-uniform,
  the error of a computation is highly dependent on its inputs.
In many cases,
  a developer must know on the range of inputs to a computation
  in order to improve its error behavior,
  but the actual intermediate inputs such computations receive
  during program execution are difficult to ascertain
  from the original, top-level program input.
\Herbgrind satisfies this need
  by computing \textit{input characteristics}
  for each symbolic expression it produces.%
\footnote{
Note that \Herbgrind's input characteristics apply
  to the inputs of symbolic expressions identified by \Herbgrind,
  not to the program inputs provided by the developer.
}
These input characteristics can show
  the ranges of each symbolic variable,
  example inputs, or other features of the expression inputs.

To compute input characteristics,
  \Herbgrind stores, for every symbolic expression,
  a summary of all values seen for that symbolic expression's free variables.
Every time a section of code (a function or a loop body, say)
  is re-executed, the inputs from that run are added to the summary.
The input characteristics system is modular,
  and \Herbgrind comes with three implementations.%
\footnote{The abstract Floatgrind analysis
  supports any arbitrary summary function on sets of input points,
  and for performance the summary function must be incremental.}
In the first kind of input characteristic,
  a representative input is selected from the input.
In the second kind of input characteristic,
  ranges are tracked for each variable in a symbolic expression.
In the third kind of input characteristic,
  ranges are tracked separately
  for positive and negative values of each variable.
\CUT{Tracking positive and negative values separately
  responds to the intuition that often values near zero
  behave quite differently from values with large magnitudes.
Future kinds of input characteristics
  could track more detailed distributional information
  or relationships between variables.}

\Herbgrind operates on a single execution of the client program
  using representative inputs provided by the developer.
During execution
  problematic code fragments typically see a range of intermediate values,
  only some of which lead to output error.%
\Herbgrind's input characteristics characterize that range of intermediate values,
  and thus rely on a well-chosen representative input for the program.
For example, in our complex plotter example in the overview,
  the function \texttt{f} is run for every pixel in the image,
  fully exercising its input range.
Since only some executions of a block of code lead to high local error,
  the input characteristics system provides two outputs
  for each characteristic and each expression:
  one for all inputs that the expression is called on,
  and one for all inputs that it has high error on.

The characteristics reported are tightly coupled
  to the symbolic expression for the relevant program fragment;
  each characteristic applies to a single variable in the expression.
For instance, when the symbolic expression is \texttt{sqrt(x+1) - sqrt(x)},
  the input characterization system might report
  that the variable \texttt{x} ranges from 1 to 1e20.
This uses the specific variable name reported in the expression,
  and applies to both nodes labeled \texttt{x} in the expression.
Since anti-unification guarantees
  that nodes assigned the same variable have taken the same values
  on each concrete expression,
  any valid summaries of the two nodes will be equivalent.

\CUT{When using \Herbgrind,
  a developer uses the input characteristics
  to guide their modifications to their code.
The developer strives to reduce error
  for inputs with characteristics similar to
  the inputs that cause high error,
  without increasing error for other inputs.}

\section{Implementation}
\label{sec:impl}

The previous section described \Herbgrind's analysis
in terms of an abstract machine; however,
important numerical software is actually written in
low level languages like C, C++, and Fortran---%
sometimes a polyglot of all three.

To support all these use cases, we refine the algorithm presented in
\Cref{sec:spec} to operate on compiled binaries instead of
abstract machine programs. \Herbgrind does this by building upon
Valgrind~\cite{valgrind}, a framework for dynamic analysis through binary
instrumentation.  Building upon Valgrind requires mapping the abstract
machine described in \Cref{sec:spec} to the VEX machine internal to
Valgrind.


Analyses built upon Valgrind receive the instructions of the client
program translated to VEX, and can add instrumentation to them freely
before they are compiled back to native machine code and executed.
\CUT{This compilation process slows down the binary,
  since VEX does not perfectly mirror machine code;
  usually the binary is about one order of magnitude slower.
VEX in many ways mirrors common machine languages
  such as \CC{x86} or \CC{ARM},
  and differs from the abstract machine model in several ways.}

Implementing the algorithm from \Cref{sec:spec} with Valgrind
  requires adapting the abstract machine's notions of
  values, storage, and operations
  to those provided by Valgrind.

\subsection{Values}
\label{ssec:values}

Unlike the abstract machine,
  VEX has values of different sizes and semantics.
Floating-point values come in different precisions,
  so the \Herbgrind implementation makes a distinction
  between single- and double-precision floating-point values.
Both types of values are shadowed with the same shadow state,
  but their behaviors in the client program are different,
  and they have different sizes in memory.
Client programs also have file descriptors, pointers,
  and integers of various sizes,
  but this does not affect \Herbgrind,
  since it does not analyze non-floating-point computations.

The algorithm in \Cref{sec:spec}
  tracks the exact value of computations
  by shadowing floating-point values with real numbers.
\Herbgrind approximates the real numbers using the MPFR library~\cite{mpfr}
  to shadow floating-point values with arbitrary-precision floating-point.%
\footnote{The precision used is configurable, set to \nDefaultPrecision by default.}
As an alternative, we could use
  an efficient library for the computable reals~\cite{computable-reals,boehm-fast,boehm-java}.%
\footnote{\Herbgrind treats real computation as an abstract data type
  and alternate strategies could easily be substituted in.}
\CUT{Other work~\cite{herbie} suggests that this precision suffices
  for capturing common floating-point problems,
  though users may need to change it for some domains.}

\subsection{Storage}
The abstract machine model of \Cref{sec:spec} represents storage as a single map
  from locations to values.
However, VEX has three different types of storage---%
  temporaries, thread state, and memory---%
  and the latter two store unstructured bytes, not values directly.
\Herbgrind uses slightly different approaches for each.

To support SIMD instructions in the SSE instruction set,
  temporaries can contain multiple floating-point values,
  unlike the memory locations in the abstract machine.
\Herbgrind attaches a \textit{shadow temporary} to each temporary:
  a type-tagged unit which can store multiple shadow values.
The shadow values stored in a shadow temporary
  correspond to the individual floating-point values
  inside a SIMD vector.
Temporaries that only store a single value
  have trivial shadow temporaries.
\CUT{SSE instructions are also often used
  for single-value floating-point operations;
  \Herbgrind handles this case properly.}

Thread state in VEX, which represents machine registers,
  is an unstructured array of bytes,
  so it does not use shadow temporaries.
Each floating-point value consumes multiple bytes,
  and floating-point values of different sizes
  take up different numbers of bytes.
This means that for reads and writes to memory,
  \Herbgrind
  must be careful to check whether they overwrite
  nearby memory locations with shadow values.
\Herbgrind also supports writing SIMD results to memory
  and reading them back at an offset,
  as long as the boundaries of individual values are respected.
In the rare cases where client programs
  make misaligned reads of floating-point values,
  \Herbgrind conservatively acts as if the read
  computes a floating-point value
  from non-floating-point inputs.

Like thread state, memory is an unstructured array of bytes,
  with the complication that it is too large to shadow completely.
\Herbgrind shadows only memory that holds floating-point values;
  memory is shadowed by a hash table from memory addresses
  to shadow values.

\CUT{\subsection{Types}

Memory and thread state in VEX is untyped, unlike in the abstract machine.
While Valgrind provides a simple type system
  to track the size of values and the semantic operations that apply to them,
  this type system is only sound for temporaries,
  and type information is lost when a value is written to thread state or memory.
Temporaries holding floating point values which are not
  immediately manipulated by floating-point operations
  are often given an integer type.
\Herbgrind runs its own static type inference within each superblock%
  \footnote{Single-entry multiple-exit sequences of instructions}
  to compute more accurate and sound type information,
  across memory and thread state reads and writes.

The \Herbgrind type system separates shadowed and unshadowed values,
  and shadowed floats of different precisions.
There are also several auxiliary types.
Locations loaded from a memory or thread state location
  which has not yet been set in the current superblock
  are given an unknown type,
  and values statically known to be non-float,
  because they came from a non-float operation,
  also have a separate type.
This type is similar to the unshadowed type,
  but will flag an error if it is used for a float operation,
  which ensures that the completeness of the analysis
  will not be silently violated.
}

\subsection{Operations}
\label{ssec:operations}

In the abstract machine model, all floating-point operations
  are handled by specialized instructions.
However, few machines support complex operations
  such as logarithms or tangents in hardware.
Instead, client programs evaluate these functions by calling libraries,
  such as the standard \CC{libm}.
\CUT{\Herbgrind can track the floating-point instructions
  used internally by these libraries,
  but they tend to be precision-specific,
  and can at best only return approximate results
  in a real-number evaluation.}
Shadowing these internal calculations directly
  would mis-compute the exact value of library calls;
\Herbgrind therefore intercepts calls to common library functions
  before building concrete expressions and measuring error.
For example, if a client program calls the \CC{tan} function,
  \Herbgrind will intercept this call
  and add \CC{tan} to the program trace,
  not the actual instructions executed by calling \CC{tan}.%
\footnote{The library wrapping system in the implementation is
  extensible: users can add a new library call to be wrapped by
  appending a single line to a python source file.}

\CUT{In Valgrind it is impossible to call the underlying implementation
  of intercepted floating-point library functions;
  \Herbgrind provides the floating-point output of the library function
  using the \CC{OpenLibm} implementation of \CC{libm}~\cite{openlibm}.}

Expert-written numerical code often uses ``compensating'' terms to
capture the error of a long chain of operations, and subtract that
error from the final result. In the real numbers, this error term
would always equal zero, since the reals don't have any error with
respect to themselves. Yet in floating point, these ``compensating''
terms are non-zero and computations that produce them therefore have
high local error. A naive implementation of \Herbgrind would therefore
report \spots influenced by every compensated operation used to compute
it, even though the compensating terms increase the accuracy of the
program.

Instead, \Herbgrind attempts to detect compensating operations,
  and not propagate influence from the compensating term
  to the output of the compensated operation.
\Herbgrind identifies compensating operations
  by looking for additions and subtractions which meet two criteria:
  they return one of their arguments when computed in the reals;
  and the output has less error than the argument which is passed through.
The influences for the other, compensating, term, are not propagated.

While most floating-point operations in real programs
  are specialized floating-point instructions or library calls,
  some programs use bitwise operations
  to implement a floating-point operation.
Programs produced by \CC{gcc} negate floating-point values
  by \CC{XOR}ing the value with a bitmap that flips the sign bit,
  and a similar trick can be used for absolute values.
\Herbgrind detects and instruments these bitwise operations,
  treating them as the operations they implement
  (including in concrete expressions).

\section{Optimization}
\label{sec:opt}

A direct implementation of the algorithm in \Cref{sec:impl}
  is prohibitively expensive.
\Herbgrind improves on it by using the classic techniques
  of laziness, sharing, incrementalization, and approximation.

\paragraph{Laziness}

Since program memory is untyped,
  it is initially impossible to tell
  which bytes in the program correspond to floating-point values.
\Herbgrind therefore tracks floating-point values in memory lazily:
  as soon as the client program executes
  a floating-point operation on bytes loaded from a memory location,
  that location is treated as a floating-point location
  and shadowed by a new shadow value.
\CUT{Some operations convert floating-point values between different formats,
  but do not manipulate the semantic content of the values themselves.
For these operations,
  \Herbgrind only shadows the computation
  if at least one of the arguments is already shadowed.
Since the output value is the same as the input value, this is sound.}

Besides lazily shadowing values in the client program,
  \Herbgrind also minimizes instrumentation.
Some thread state locations can always be ignored,
  such as CPU flag registers.
VEX also adds a preamble to each basic block
  representing the control flow effects of the architecture,
  which \Herbgrind also ignores.

For more fine-grained instrumentation minimization,
  \Herbgrind makes use of static superblock type analysis.
Values known to be integers do not have to be instrumented,
  and values known to be floating point can have type checking elided.
This combination of static type analysis and dynamic error analysis
  is crucial for reducing \Herbgrind's overhead.
Unfortunately, \Herbgrind must still instrument
  many simple memory operations,
  since values that are between storage locations but not operated on
  could have shadow values.

The static type analysis is also used
  to reduce reduce calls from the instrumentation
  into \Herbgrind C functions.
Valgrind allows the instrumentation
  to call into C functions provided by \Herbgrind, which then
  compute shadow values, build concrete expressions, and track influences.
However, calls from client program to host functions are slow.
The static type analysis allows inlining these computations
  directly into VEX, avoiding a client-host context switch,
  because the type system tracks the size
  of values in thread state.
Knowing the size means no type or size tests need to be done,
  so instrumentation can be inlined
  without requiring branches and thus crossing superblock boundaries.
Inlining is also used for copies between temporaries,
  and for some memory accesses,
  where the inlined code must also update reference counts.

\paragraph{Sharing}


Many floating-point values are copies of each other,
  scattered in temporaries, thread state, and memory.
Though copying floating-point values is cheap on most architectures,
  copying shadow values requires copying
  MPFR values, concrete expressions, and influence sets.
To save time and memory,
  shadow values are shared between copies.
Shadow values are reference counted to ensure that they can be discarded
  once they no longer shadow any floating-point values.
The trace nodes stored in a shadow value
  are not freed along with the shadow value,
  since traces also share structure.
Traces are therefore reference counted as well,
  with each shadow value holding a reference to its trace node,
  and each trace node holding references to its children.

Many shadow values are freed shortly after they are created.
Related data structures, like trace nodes,
 are also allocated and freed rapidly,
 so memory allocation quickly becomes a bottleneck.
\Herbgrind uses custom stack-backed pool allocators
  to quickly allocate and free many objects of the same size.

\paragraph{Incrementalization}

The algorithm in \Cref{sec:spec}
  accumulates errors, concrete expressions, and operation inputs per-instruction,
  and summarizes all the results after the program finishes running.
For long-running programs,
  this approach requires storing large numbers of
  ever-growing concrete expressions.
The implementation of \Herbgrind avoids this problem
  by aggregating
  errors (into average- and maximum- total and local errors)
  concrete expressions (into symbolic expressions)
  and inputs (into input characteristics)
  incrementally, as the analysis runs.
This leads to both large memory savings and significant speed-ups.

This incrementalization does not change the analysis results
  since our implementation of anti-unification,
  used to aggregate concrete expressions,
  and summation, used to aggregate error,
  are associative.
\CUT{Storing only symbolic expressions
  also allows freeing more concrete expressions,
  leading to lower memory usage and therefore faster runtime.}

\subsection{Approximation}
\label{ssec:approximation}

\Herbgrind makes a sound, but potentially incomplete, approximation
  to the standard anti-unification algorithm to speed it up.
Anti-unification requires knowing which pairs of nodes are equivalent,
  so that those nodes could be generalized to the same variable.
Symbolic expressions can be trees hundreds of nodes deep,
  and this equivalence information must be recomputed
  at every node,
  so computing these equivalence classes for large trees
  is a significant portion of \Herbgrind's runtime.
To limit the cost,
  \Herbgrind exactly computes equivalence information
  to only a bounded depth for each node,
  \nDepthBound by default.
In practice, this depth suffices
  to produce high-quality symbolic expressions.
This depth bound also allows
  freeing more concrete program trace nodes,
  further reducing memory usage.

\section{Case Studies}
\label{sec:cases}

This section describes three examples
  where \Herbgrind was used to identify the root causes of floating-point errors
  in numerical programs:
  in all three cases, the bugs were fixed in later versions of the software.
\Herbgrind's three major subsystems were crucial in detecting and understanding these bugs.

\paragraph{Gram-Schmidt Orthonormalization}

Gram-Schmidt orthonormalization transforms a collection of vectors
  into a orthogonal basis of unit vectors for their span.
We investigated an implementation of Gram-Schmidt orthonormalization
  provided by the Polybench benchmark suite for numerical kernels.
Polybench is provided in several languages;
  we used the C version of Polybench~3.2.1\@.

In the Gram-Schmidt orthonormalization kernel,
  \Herbgrind detected a floating-point problem
  which it reported to have 64 bits of error,
  surprising for a vetted numerical benchmark.
Upon investigating,
  we found that Gram-Schmidt decomposition
  is not well defined on the given inputs, resulting in a division by zero;
  \Herbgrind reports the resulting \CC{NaN} value as having maximal error.
The fundamental problem is not in the Gram-Schmidt procedure itself
  but in its invocation on an invalid intermediate value.
Luckily, \Herbgrind provides,
  as its example problematic input to the computation in question,
  a zero vector, an invalid input to Gram-Schmidt orthonormalization.
\Herbgrind's input characteristics system
  was able to link the error in the program output
  to the root cause of an input violating the precondition
  of the orthnormalization procedure.
Note that there was nothing wrong with the procedure itself,
  but rather its interaction with the program around it.
Upon understanding the bug (and fixing it ourselves),
  we tested version~4.2.0 of Polybench
  and confirmed that this more recent version fixed the problem
  by changing the procedure that generated the vectors
  to ensure that valid inputs to Gram-Schmidt orthonormalization are produced.

\paragraph{PID Controller}
A proportional-integral-derivative controller
  is a control mechanism widely used in industrial control systems.
The controller attempts to keep
  some \textit{measure} at a fixed value.
It runs in a loop, receiving the current value of the measure as input
  and outputting the rate at which to increase or decrease the measure.
We investigated an adaptation of a simple PID controller
  which runs for a fixed number of iterations
  and with a fixed rate of change to the measure~\cite{damouche-martel-chapoutot-nsv14}.

We initially ran \Herbgrind on the PID controller
  expecting to find, perhaps, some floating-point error
  in the controller code itself.
Instead, we found that \Herbgrind was detecting a problem
  in the loop condition.
To run the PID controller for a limited number of seconds,
  the program tests the condition \texttt{(t < N)},
  where N is the number of seconds.
The variable \texttt{t} is stored as a double-precision
  floating-point number, and is incremented by \texttt{0.2}
  on every iteration through the loop.
As we experimented with different loop bounds,
  \Herbgrind noticed that the condition,
  for some loop bounds, iterates once too many times.
For example, if the loop bound is set to \texttt{10.0},
  the loop executes 51 times, not 50 times,
  because adding \texttt{0.2} to itself 50 times
  produces a value $3.5\cdot10^{-15}$ less than 10.
This bug is closely related to one that occurred
  in the Patriot missile defense system in 1992,
  resulting in the death of 28 people~\cite{patriot}.
\Herbgrind's automatic marking of all control flow operations as spots
  was necessary to detect the bug and link the inaccurate increment
  to its affect on control flow.
\Herbgrind was successfully able to trace back
  from error detected in the output of the program
  to the \emph{root cause} of the error,
  the inaccurate increment;
  the output contained the source location of the erroneous compare
  and reported that it was influenced by the inaccurate increment.
We notified the authors of the adapted PID controller
  and they confirmed the bug
  and identified a fix: incrementing the \texttt{t} variable
  by 1 instead of 0.2, and changing the test to \texttt{(t * 0.2 < N)}.

\paragraph{\textsc{Gromacs}}

\textsc{Gromacs} is a molecular dynamics package
  used for simulating proteins, lipids, and nucleic acids
  in drug discovery, biochemistry, and molecular biology.
We investigated the version of \textsc{Gromacs}
  that ships with the SPEC CPU 2006 benchmark suite.
\textsc{Gromacs} is a large program---%
  42\thinspace762 lines of C, with the inner loops,
  which consume approximately 95\% of the runtime, written in 21\thinspace824 lines of Fortran.
We tested \textsc{Gromacs} on the \CC{test} workload
  provided by CPU~2006, which simulates the protein Lysozyme in a water-ion solution.

During this run, \Herbgrind reported an error
  in the routine that computes dihedral angles
  (the angle between two planes, measured in a third, mutual orthogonal plane).
For inputs where the dihedral angle is close to flat,
  corresponding to four colinear molecules,
  the dihedral angle computation was returning values
  with significant error due to cancellation
  in the computation of a determinant.
These cases, though a small subset of all possible angles, were important.
First, collections of four colinear molecules are common,
  for example in triple-bonded organic compounds such as alkynes.
Second, molecular dynamics is chaotic,
  so even small errors can quickly cause dramatically different behavior.

\Herbgrind's symbolic expression system
  was crucial in understanding the root cause of this bug.
The dihedral angle procedure invokes code from multiple source files,
  across both C and Fortran, moving data into and out of vector data structures.
The symbolic expression gathered together the slivers of computation
  that contributed to the high rounding error.
From the expression reported by \Herbgrind the potential for cancellation was clear
  and the input characteristics provided by \Herbgrind
  allowed us to narrow our investigations to flat angles.
We identified the problem and developed a fix
  based on the numerical analysis literature~\cite{tetgen}.
After we contacted the developers,
  they confirmed the bug and explained that they had deployed
  a similar fix in recent \textsc{Gromacs} versions.

\section{Evaluation}
\label{sec:eval}

This section shows that \Herbgrind identifies correct root causes of error
  in inaccurate floating-point program binaries,
  and that the root causes are reported with sufficient precision
  to allow improving accuracy.
The first subsection demonstrates this
  for \Herbgrind in its default configuration,
  while the second subsection examines the effect
  of \Herbgrind's various tunable parameters.
Each experiment uses the standard FPBench suite
  of general-purpose floating-point programs~\cite{fpbench}.

\subsection{Improvability}

The true root cause of a floating-point inaccuracy
  is a part of the inaccurate computation
  which can be rewritten to reduce error.
\Herbgrind's value is its ability to find true root causes;
  thus, this evaluation measures the fraction
  of true root causes found by \Herbgrind,
  and the fraction of \Herbgrind's candidate root causes
  that are true root causes.

\paragraph{Methodology}

To determine whether a candidate is a true root cause,
  one must determine whether its error can be improved,
  which depends on the expertise of the programmer.
As a mechanical, easily quantifiable proxy,
  we picked a state-of-the-art tool, Herbie~\cite{herbie},
  and used it to determine which of \Herbgrind's identified root causes
  were improvable.
%
To use \Herbgrind on the FPBench benchmarks,
  these benchmarks must be compiled to native code.
We do so by using the publicly available
FPCore-to-C compiler provided with FPBench, and then compiling this C
code, along with some driver code which exercises the benchmarks on
many inputs, using the GNU C Compiler. We then run the binary under
\Herbgrind, and pass the resulting output to Herbie.
We also timed the benchmarks to measure their speed.%
\footnote{Our original timing code actually had a floating point bug,
  which we discovered when \Herbgrind included it in its output.}

All experiments were run on an Intel Core i7-4790K processor at 4GHz
with 8 cores, running Debian 9 with 32 Gigabytes of memory.%
\footnote{Results were obtained using GNU Parallel~\cite{gnu-parallel}}.
\Herbgrind introduces a 574x overhead on the FPBench suite.

Not every benchmark in the FPBench suite exhibits significant error,
  nor can Herbie improve all of the benchmarks that exhibit error.
To provide a valid comparison for \Herbgrind's results,
  we compare \Herbgrind against an ``oracle''
  which directly extracts the relevant symbolic expression from source benchmark.

\paragraph{Results}

The oracle finds that, of 86 benchmarks,
  30 have significant error (> 5 bits).
Of these, 29 are determined by \Herbgrind
  to have significant error.

Of the 30 benchmarks with significant error,
  the oracle produces an improvable root cause for all 30 benchmarks.

\Herbgrind determines candidate root causes
  for 29 of the errors (96\%), and for 25 of the benchmarks,
  Herbie detects significant error in
  the candidate root causes reported (86\%).
The remaining programs reflects
  either limitations in \Herbgrind's ability
  to properly identify candidate root causes%
\footnote{
  Candidate root causes in which Herbie can not independently detect error
  mostly reflect limitations in \Herbgrind's ability to accurately characterize inputs.
  Note that the input characterization system is modular and easily extended.},
  or limitations in Herbie's ability to sample inputs effectively.
Finally, of the 30 total benchmarks which had error
  detectable by the oracle and Herbie,
  \Herbgrind can produce improvable root causes for 25 (83\%).

Overall, \Herbgrind is able to determine the true root cause
  for 25 of the programs in the FPBench suite,
  demonstrating that it is useful for diagnosing and fixing
  floating-point inaccuracies.

\subsection{Subsystems}
\Herbgrind's analysis consists of three main subsystems, influence
tracking, symbolic expressions, and input characteristics. \CUT{Each
  of these systems have main parameters that control their behavior.}
In this section we will demonstrate the effect of these subsystems.

\begin{figure*}
  \begin{subfigure}[t]{.24\linewidth}
    \includegraphics[width=\linewidth]{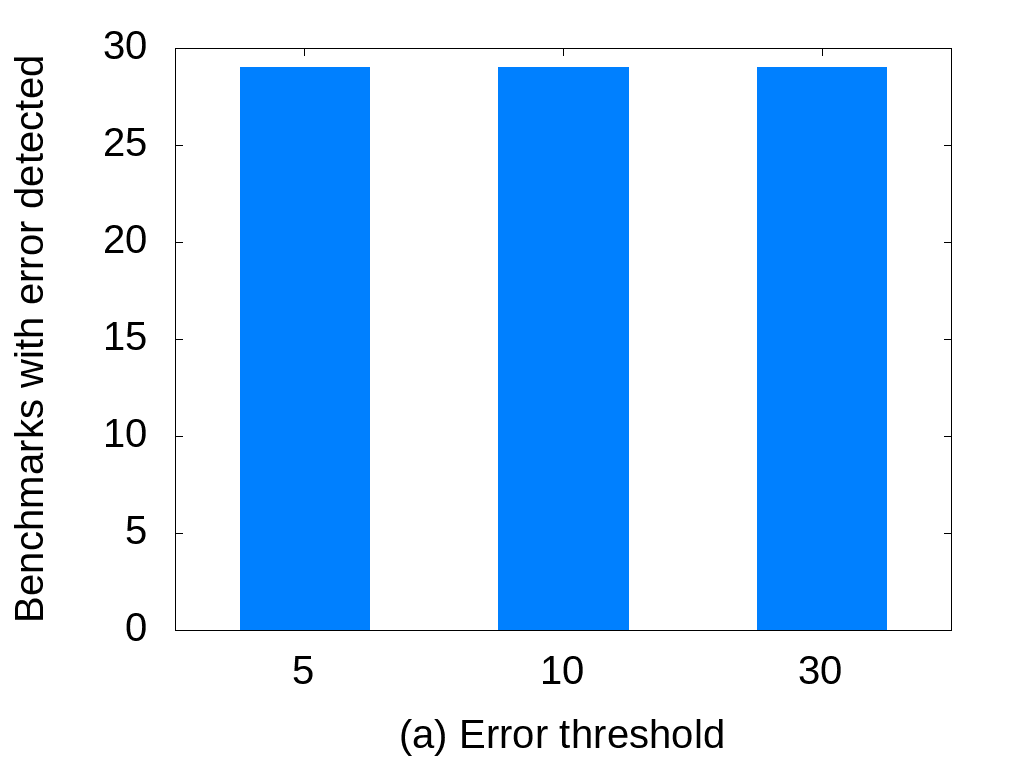}
  \end{subfigure}%
  \hfill%
  \begin{subfigure}[t]{.24\linewidth}
    \includegraphics[width=\linewidth]{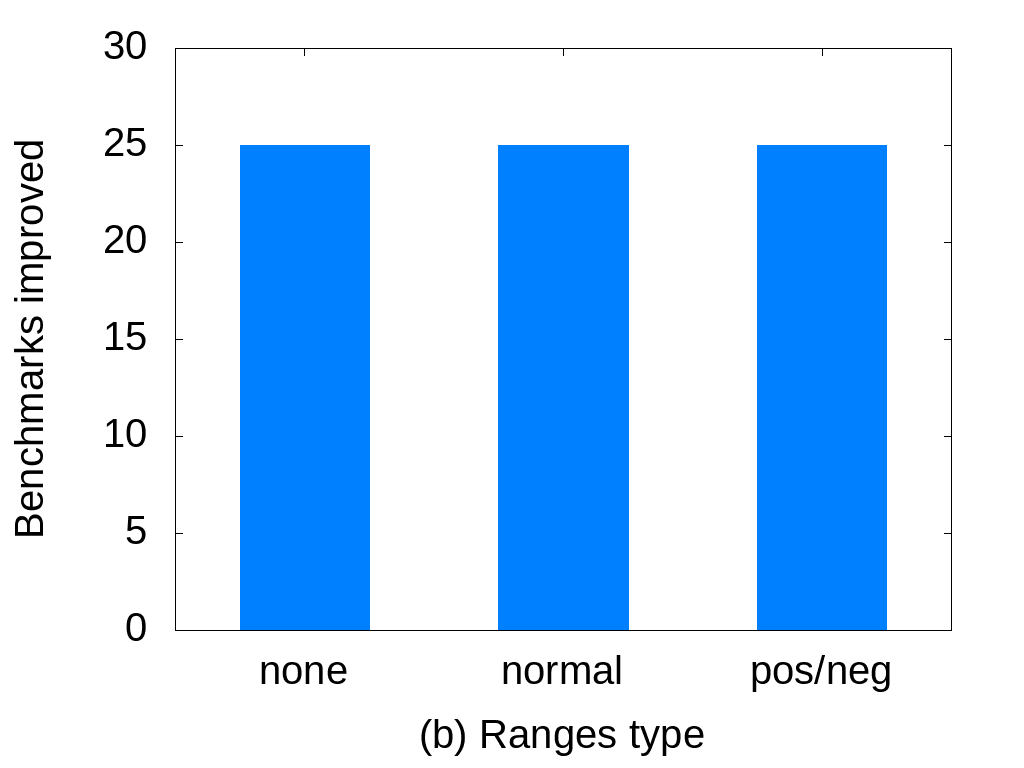}
  \end{subfigure}%
  \hfill%
  \begin{subfigure}[t]{.24\linewidth}
    \includegraphics[width=\linewidth]{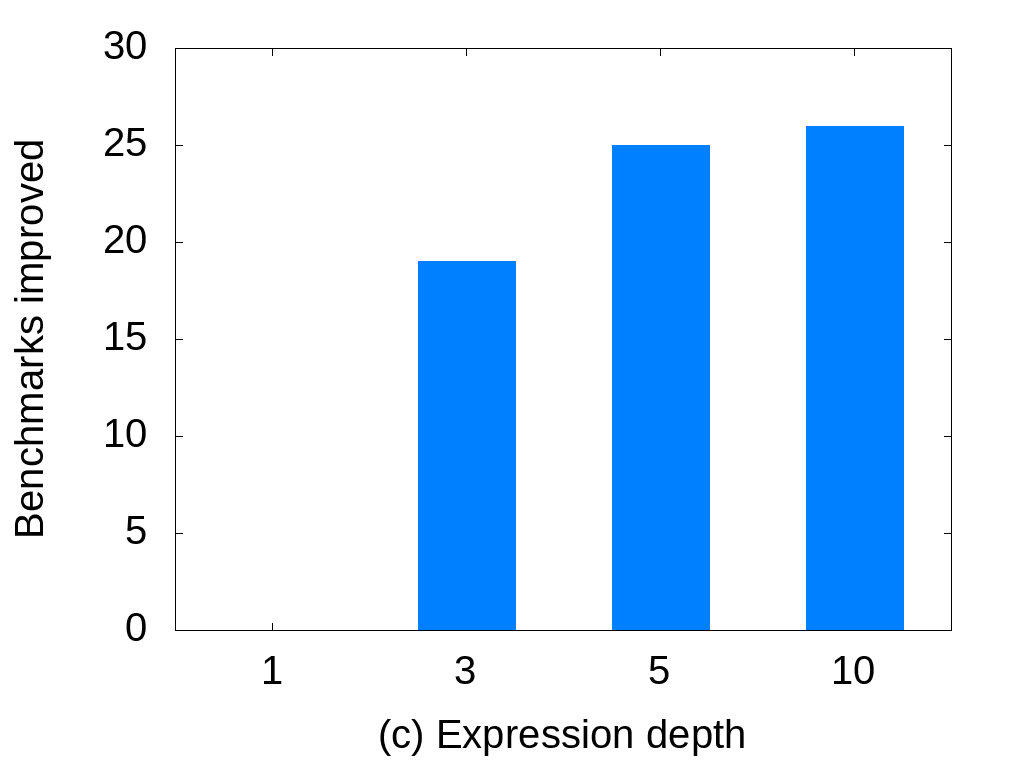}
  \end{subfigure}
  \hfill%
  \begin{subfigure}[t]{.24\linewidth}
    \includegraphics[width=\linewidth]{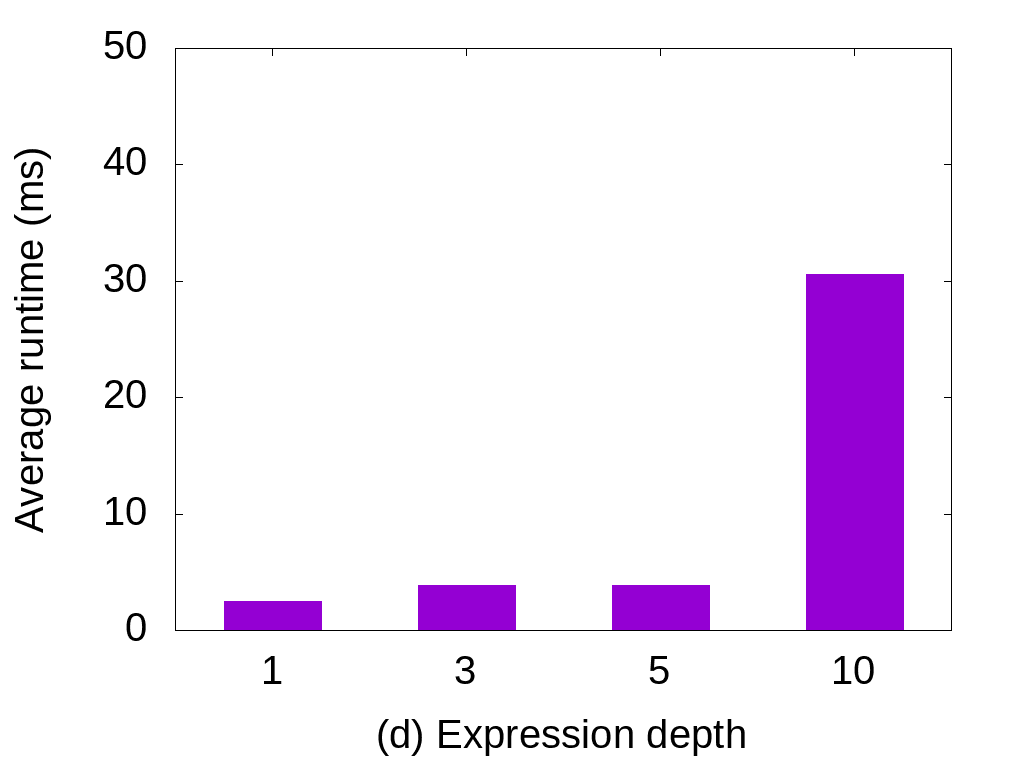}
  \end{subfigure}%
  \caption{In (a) we compare the number of computations flagged with
    various error thresholds. In (b) we show how many benchmarks can
    be improved with various types of ranges. (c) and (d) explore the
    effect of different maximum expression depths on runtime and
    number of benchmarks improved.}
  \label{fig:eval-graphs}
\end{figure*}

\Cref{fig:eval-graphs}a shows the result of running
\Herbgrind with various error thresholds for the influences
system. The error threshold selected determines how much local error
an expression has to have before it is marked as ``significantly
erroneous'', and tracked as a source of error. \CUT{For this experiment, we
used error thresholds of 5-, 10-, and 30-bits of error.}

A higher threshold means that fewer computations are reported as being
problematic. Users might want to use a higher error threshold on
certain applications when there are too many expressions that are
somewhat erroneous to address them all. \CUT{Raising the error
  threshold in this case makes it easier to prioritize certain sources
  of error.} In highly critical applications, where it is important to
address even a small source of error, users might choose to lower the
error threshold to catch even more errors. \CUT{The error threshold
  used has no significant impact on the runtime of the analysis.}

\CUT{The next subsystem of \Herbgrind that users might want to test is
  the symbolic expressions system.} Because of the non-local nature of
floating-point error, the root cause of error is often spread across
many floating-point operations. To measure how far root causes are
spread, we varied the maximum expression depth \Herbgrind
tracked. \CUT{Expressions beyond this depth are generalized until they
  are within this depth.} In \Cref{fig:eval-graphs}c and
\Cref{fig:eval-graphs}d, we measured the runtime and
effectiveness of \Herbgrind using various maximum expression
depths.\CUT{ of 1, 3, and 5 nodes deep.}

A maximum expression depth of 1 node deep effectively disables
symbolic expression tracking, and only reports the operation where
error is detected, much like FpDebug and similar floating-point
debugging tools. However, unlike those tools, it still tracks the
influence of error and the range of inputs to the operation. As you
can see from the figure, not tracking operations before the one that
produced error results in a speedup over the normal configuration, but
at a high cost: none of the expressions produced are significantly
improvable.
\CUT{A maximum expression depth of 3 nodes deep does significantly better,
but still does not capture the complete root causes for all the
benchmarks. An expression depth of 5 is the default, and produces the
number of influences listed above.}

Finally, to test the effectiveness of input characterization, we
measured the improvability of our benchmarks in three configurations:
with ranges turned off \CUT{(no precondition produced)}, with a single
range for all inputs, and with separate ranges for positive and
negative inputs\CUT{In this case, the default configuration was not
  the most detailed one, but was in fact the total ranges
  configuration; positive and negative ranges are more fine-grained,
  but are not default.}
(see \Cref{fig:eval-graphs}b). In this dataset it appears that input
ranges do not significantly affect results; however, this could be due
to the fact that these programs are small micro-benchmarks.

\CUT{\paragraph{Performance}

The strength of \Herbgrind's root cause analysis comes at a cost: on
average it slows our benchmarks down by 574x. \Herbgrind's
analysis is not suitable for running on a full production dataset, but
instead should be run on a small test dataset that exercises all
aspects of the code. As a debugging tool, \Herbgrind should never be
used in a production run of software: even the minimal overhead of the
instrumentation framework is too much for many applications in
production. Instead, \Herbgrind is intended to be used to find and fix
bugs \textit{before} production runs.

As a system with a few similar features,
  FpDebug seems a useful point of comparison for the performance of \Herbgrind.
Unfortunately we were unable to find a working implementation
  of FpDebug to run on our benchmarks.
\Cref{fig:feature-table} compares the \Herbgrind's performance
  to the slowdown numbers published in the FpDebug paper~\cite{fpdebug}.
\CUT{
Instead, we built a modified version of \Herbgrind which
emulates the behavior of FpDebug, providing a error detector through
high-precision shadow values, but nothing else~\footnote{Unlike
  FpDebug, \Herbgrind's implementation of high-precision shadows still
  includes SIMD support, bit-level operations, and library
  wrapping}. We found that \Herbgrind's root cause analysis has a
slowdown of \todo{number} over a simple error detector.
}
}

\paragraph{Library Wrapping}

\Herbgrind instruments calls to mathematical library functions
  such as \CC{sqrt} and \CC{tan}
  to correctly evaluate the exact result of a computation
  and provide simpler symbolic expressions.
With this wrapping behavior turned off,
  \Herbgrind finds significantly more complex expressions,
  representing the internals of library functions:
  the largest expressions are not 9 but 31 operations in size,
  and 133 expressions%
\footnote{
  With library wrapping disabled, \Herbgrind identifies
  848 problematic expressions, mostly corresponding
  to false positives in the internals of the math library.}
  have more than 9 operations.
For example, instead of $e^x - 1$,
  \Herbgrind finds 17 expressions such as
\[
(x - 0.6931472(y - 6.755399\mathtt{e}15) + 2.576980\mathtt{e}10) - 2.576980\mathtt{e}10.
\]
Furthermore, as discussed in \Cref{sec:impl}, without wrapping
  calls to mathematical libraries, \Herbgrind measures output accuracy incorrectly,
  though on the FPBench examples the inaccuracy is slight.

\subsection{Handling Expert Tricks}

Detecting compensating terms (see \Cref{ssec:operations}) in client code is
important for reducing false positives in \Herbgrind's output. To test the
compensation detection system, we applied \Herbgrind to analyze Triangle, an
expert-written numerical program.

Triangle~\cite{triangle}, written in C by Jonathan Shewchuk,
  is a mesh generation tool,
  which computes the Delaunay triangulation of a set of input points,
  and can add additional points so that
  the triangulation produced satisfies various stability properties,
  such as avoiding particularly sharp angles.
Running on Triangle's example inputs,
  we found that \Herbgrind's compensation detection
  correctly handles all but 14 of 225 compensating terms with local error
  and does not present these false positives to the user.

The 14 remaining compensated operations
  are not detected, because the compensating term affects control flow:
  Triangle checks whether the compensating term is too large,
  and if so runs the same computation in a different way.
\Herbgrind's real-number execution computes the accurate value
  of a compensating term to be 0,
  so these branches often go the ``wrong way''.
Fortunately, given \Herbgrind's candidate root cause,
  this behavior is always easy to check in the Triangle source.

\section{Related Work}
\label{sec:related}

There is a rich literature on analyzing and mitigating floating-point error.
Below we discuss the most closely related work.


Recently, work on statically analyzing error for floating point programs
has made tremendous progress~\cite{higham-book,salsa-1,salsa-2,fluctuat,rosa,fptaylor,fptuner}.
Broadly, this work focuses on providing sound, though conservative,
  error bounds for small numerical kernels.
This work is useful for reasoning about the expressions returned by \Herbgrind,
  but does not on its own scale to large numerical software.

Several papers in recent years have used dynamic analyses
  to analyze floating-point error.

FpDebug~\cite{fpdebug} uses Valgrind
  to build a dynamic analysis of floating point error.
Like \Herbgrind, it uses MPFR shadow values
  to measure the error of individual computations.
Unlike FpDebug however, \Herbgrind's shadow real execution
  is based on a model of full programs,
  including control flow, conversions, and I/O,
  as opposed to FpDebug's model of VEX blocks.
This enables a rigorous treatment of branches as spots,
  and leads to extensions such as wrapping library functions,
  sharing shadow values, SIMD operations, and bit-level transformations.
All these features required significant design and engineering
  to scale to 300 KLOC numerical benchmarks from actual scientific computing applications.
In addition to an improved real execution,
  \Herbgrind departs from FpDebug
  with its \spots and influences system, symbolic expressions, and input ranges,
  which allow it to connect inaccurate floating-point expressions
  to inaccurate outputs produced by the program.
\Herbgrind's use of local error, symbolic expressions, and input ranges,
  help the user diagnose the parts of the program
  that contributed to the detected error.

Similarly to FpDebug and \Herbgrind,
  Verrou~\cite{verrou} is a dynamic floating point analysis built on Valgrind.
Verrou's aim is also to detect floating point error in numerical software,
  but attempts to do so at much lower overhead.
The resulting approach uses very little instrumentation
  to perturb the rounding of a floating point program,
  thus producing a much more conservative report
  of possible rounding errors.

Recent work by Bao and Zhang~\cite{baozhang} also attempts to detect floating-point
error with low overhead, with the goal of determining when
floating-point error flows into what they call ``discrete
factors''. The tool is designed to detect the possible presence of
inaccuracy with very low runtime overhead to enable re-running in
higher precision. In this context, a very high false positive rate (>
80-90\% in their paper) is acceptable, but it is not generally
acceptable as a debugging technique. Bao and Zhang's discrete factors
address only floating-point errors that cause changes in integer or
boolean values (hence ``discrete''). Unlike~\cite{baozhang},
\Herbgrind tracks all factors (not just discrete ones), including
changes in floating-point values that lead to changes in
floating-point outputs. Re-running programs in higher precision is
untenable in many contexts, but may work for some.

Herbie~\cite{herbie} is a tool
  for the automatically improving the accuracy
  of small floating point expressions ($\approx$ 10 LOC).
Herbie uses randomly sampled input points
  and an MPFR-based ground truth
  to evaluate expression error.
This statistical, dynamic approach to error
  cannot give sound guarantees,
  but is useful for guiding a search process.
Herbie's main focus
  is on suggesting more-accurate floating-point expressions
  to program developers.
Herbie can be combined with \Herbgrind
  to improve problematic floating point code
  in large numerical programs,
  by feeding the expressions produced by \Herbgrind
  directly into Herbie to improve them.

Wang, Zou, He, Xiong, Zhang, and Huang~\cite{fse16}
  develop a heuristic to determine which instructions in
  core mathematical libraries
  have an implicit dependence on the precision of the floating-point numbers.
A ground truth for such precision-specific operations
  cannot be found by evaluating the operations at higher precision.
These results justify \Herbgrind
  detecting and abstracting calls to the floating-point math library.

CGRS~\cite{cgrs} uses evolutionary search to find inputs
  that cause high floating-point error;
  these inputs can be used for debugging or verification.
Unlike \Herbgrind, these inputs can be unrealistic
  for the program domain,
  and CGRS does not help the developer
  determine which program expressions created the high error.
However, users who want
  to analyze the behavior of their programs on such inputs
  can use \Herbgrind to do so.

\section{Conclusion}
\label{sec:conclusion}

Floating point plays a critical role in applications supporting
science, engineering, medicine, and finance. This paper presented
\Herbgrind, the first approach to identifying candidate root causes of
floating point errors in such software. \Herbgrind does this with
three major subsystems: a shadow taint analysis which tracks the
influence of error on important program locations, a shadow symbolic
execution which records the computations that produced each value, and
an input characterization system which reports the inputs to
problematic computations. \Herbgrind's analysis is implemented on top
of the Valgrind framework, and finds bugs in standard numerical
benchmarks and large numerical software written by experts.



\citestyle{acmnumeric}   


\begin{thebibliography}{38}


\ifx \showCODEN    \undefined \def \showCODEN     #1{\unskip}     \fi
\ifx \showDOI      \undefined \def \showDOI       #1{#1}\fi
\ifx \showISBNx    \undefined \def \showISBNx     #1{\unskip}     \fi
\ifx \showISBNxiii \undefined \def \showISBNxiii  #1{\unskip}     \fi
\ifx \showISSN     \undefined \def \showISSN      #1{\unskip}     \fi
\ifx \showLCCN     \undefined \def \showLCCN      #1{\unskip}     \fi
\ifx \shownote     \undefined \def \shownote      #1{#1}          \fi
\ifx \showarticletitle \undefined \def \showarticletitle #1{#1}   \fi
\ifx \showURL      \undefined \def \showURL       {\relax}        \fi
\providecommand\bibfield[2]{#2}
\providecommand\bibinfo[2]{#2}
\providecommand\natexlab[1]{#1}
\providecommand\showeprint[2][]{arXiv:#2}

\bibitem[\protect\citeauthoryear{Altman, Gill, and McDonald}{Altman
  et~al\mbox{.}}{2003}]%
        {num-issues-in-stat}
\bibfield{author}{\bibinfo{person}{Micah Altman}, \bibinfo{person}{Jeff Gill},
  {and} \bibinfo{person}{Michael~P. McDonald}.}
  \bibinfo{year}{2003}\natexlab{}.
\newblock \bibinfo{booktitle}{{\em Numerical Issues in Statistical Computing
  for the Social Scientist}}.
\newblock \bibinfo{publisher}{Springer-Verlag}. 1--11 pages.
\newblock


\bibitem[\protect\citeauthoryear{Altman and McDonald}{Altman and
  McDonald}{2003}]%
        {num-replication}
\bibfield{author}{\bibinfo{person}{Micah Altman} {and}
  \bibinfo{person}{Michael~P. McDonald}.} \bibinfo{year}{2003}\natexlab{}.
\newblock \showarticletitle{Replication with attention to numerical accuracy}.
\newblock \bibinfo{journal}{{\em Political Analysis\/}} \bibinfo{volume}{11},
  \bibinfo{number}{3} (\bibinfo{year}{2003}), \bibinfo{pages}{302--307}.
\newblock
\showURL{%
\url{http://pan.oxfordjournals.org/content/11/3/302.abstract}}


\bibitem[\protect\citeauthoryear{Bao and Zhang}{Bao and Zhang}{2013}]%
        {baozhang}
\bibfield{author}{\bibinfo{person}{Tao Bao} {and} \bibinfo{person}{Xiangyu
  Zhang}.} \bibinfo{year}{2013}\natexlab{}.
\newblock \showarticletitle{On-the-fly Detection of Instability Problems in
  Floating-point Program Execution}.
\newblock \bibinfo{journal}{{\em SIGPLAN Not.\/}} \bibinfo{volume}{48},
  \bibinfo{number}{10} (\bibinfo{date}{Oct.} \bibinfo{year}{2013}),
  \bibinfo{pages}{817--832}.
\newblock
\showISSN{0362-1340}
\showDOI{%
\url{https://doi.org/10.1145/2544173.2509526}}


\bibitem[\protect\citeauthoryear{Benz, Hildebrandt, and Hack}{Benz
  et~al\mbox{.}}{2012}]%
        {fpdebug}
\bibfield{author}{\bibinfo{person}{Florian Benz}, \bibinfo{person}{Andreas
  Hildebrandt}, {and} \bibinfo{person}{Sebastian Hack}.}
  \bibinfo{year}{2012}\natexlab{}.
\newblock \showarticletitle{A Dynamic Program Analysis to Find Floating-point
  Accuracy Problems} {\em (\bibinfo{series}{PLDI '12})}.
  \bibinfo{publisher}{ACM}, \bibinfo{address}{New York, NY, USA},
  \bibinfo{pages}{453--462}.
\newblock
\showISBNx{978-1-4503-1205-9}
\showURL{%
\url{http://doi.acm.org/10.1145/2254064.2254118}}


\bibitem[\protect\citeauthoryear{Boehm}{Boehm}{2004}]%
        {boehm-java}
\bibfield{author}{\bibinfo{person}{Hans-J. Boehm}.}
  \bibinfo{year}{2004}\natexlab{}.
\newblock \showarticletitle{The constructive reals as a Java Library}.
\newblock \bibinfo{journal}{{\em J. Log. Algebr. Program\/}}
  \bibinfo{volume}{64} (\bibinfo{year}{2004}), \bibinfo{pages}{3--11}.
\newblock


\bibitem[\protect\citeauthoryear{Chiang, Baranowski, Briggs, Solovyev,
  Gopalakrishnan, and Rakamari\'{c}}{Chiang et~al\mbox{.}}{2017}]%
        {fptuner}
\bibfield{author}{\bibinfo{person}{Wei-Fan Chiang}, \bibinfo{person}{Mark
  Baranowski}, \bibinfo{person}{Ian Briggs}, \bibinfo{person}{Alexey Solovyev},
  \bibinfo{person}{Ganesh Gopalakrishnan}, {and} \bibinfo{person}{Zvonimir
  Rakamari\'{c}}.} \bibinfo{year}{2017}\natexlab{}.
\newblock \showarticletitle{Rigorous Floating-point Mixed-precision Tuning}. In
  \bibinfo{booktitle}{{\em Proceedings of the 44th ACM SIGPLAN Symposium on
  Principles of Programming Languages}} {\em (\bibinfo{series}{POPL 2017})}.
  \bibinfo{publisher}{ACM}, \bibinfo{address}{New York, NY, USA},
  \bibinfo{pages}{300--315}.
\newblock
\showISBNx{978-1-4503-4660-3}
\showDOI{%
\url{https://doi.org/10.1145/3009837.3009846}}


\bibitem[\protect\citeauthoryear{Chiang, Gopalakrishnan, Rakamari\'c, and
  Solovyev}{Chiang et~al\mbox{.}}{2014}]%
        {cgrs}
\bibfield{author}{\bibinfo{person}{Wei-Fan Chiang}, \bibinfo{person}{Ganesh
  Gopalakrishnan}, \bibinfo{person}{Zvonimir Rakamari\'c}, {and}
  \bibinfo{person}{Alexey Solovyev}.} \bibinfo{year}{2014}\natexlab{}.
\newblock \showarticletitle{Efficient Search for Inputs Causing High
  Floating-point Errors}. \bibinfo{publisher}{ACM}, \bibinfo{pages}{43--52}.
\newblock


\bibitem[\protect\citeauthoryear{Damouche, Martel, and Chapoutot}{Damouche
  et~al\mbox{.}}{2015a}]%
        {salsa-2}
\bibfield{author}{\bibinfo{person}{Nasrine Damouche}, \bibinfo{person}{Matthieu
  Martel}, {and} \bibinfo{person}{Alexandre Chapoutot}.}
  \bibinfo{year}{2015}\natexlab{a}.
\newblock \showarticletitle{Formal Methods for Industrial Critical Systems:
  20th International Workshop, FMICS 2015 Oslo, Norway, June 22-23, 2015
  Proceedings}.
\newblock  (\bibinfo{year}{2015}), \bibinfo{pages}{31--46}.
\newblock
\showISBNx{978-3-319-19458-5}


\bibitem[\protect\citeauthoryear{Damouche, Martel, and Chapoutot}{Damouche
  et~al\mbox{.}}{2015b}]%
        {damouche-martel-chapoutot-nsv14}
\bibfield{author}{\bibinfo{person}{N. Damouche}, \bibinfo{person}{M. Martel},
  {and} \bibinfo{person}{A. Chapoutot}.} \bibinfo{year}{2015}\natexlab{b}.
\newblock \showarticletitle{Transformation of a \{PID\} Controller for
  Numerical Accuracy}.
\newblock \bibinfo{journal}{{\em Electronic Notes in Theoretical Computer
  Science\/}}  \bibinfo{volume}{317} (\bibinfo{year}{2015}), \bibinfo{pages}{47
  -- 54}.
\newblock
\showISSN{1571-0661}
\showDOI{%
\url{https://doi.org/10.1016/j.entcs.2015.10.006}}
\newblock
\shownote{The Seventh and Eighth International Workshops on Numerical Software
  Verification (NSV).}


\bibitem[\protect\citeauthoryear{Damouche, Martel, Panchekha, Qiu,
  Sanchez-Stern, and Tatlock}{Damouche et~al\mbox{.}}{2016}]%
        {fpbench}
\bibfield{author}{\bibinfo{person}{Nasrine Damouche}, \bibinfo{person}{Matthieu
  Martel}, \bibinfo{person}{Pavel Panchekha}, \bibinfo{person}{Jason Qiu},
  \bibinfo{person}{Alex Sanchez-Stern}, {and} \bibinfo{person}{Zachary
  Tatlock}.} \bibinfo{year}{2016}\natexlab{}.
\newblock \showarticletitle{Toward a Standard Benchmark Format and Suite for
  Floating-Point Analysis}.
\newblock  (\bibinfo{date}{July} \bibinfo{year}{2016}).
\newblock


\bibitem[\protect\citeauthoryear{Darulova and Kuncak}{Darulova and
  Kuncak}{2014}]%
        {rosa}
\bibfield{author}{\bibinfo{person}{Eva Darulova} {and} \bibinfo{person}{Viktor
  Kuncak}.} \bibinfo{year}{2014}\natexlab{}.
\newblock \showarticletitle{Sound Compilation of Reals} {\em
  (\bibinfo{series}{POPL '14})}. \bibinfo{publisher}{ACM},
  \bibinfo{address}{New York, NY, USA}, \bibinfo{pages}{235--248}.
\newblock
\showISBNx{978-1-4503-2544-8}
\showURL{%
\url{http://doi.acm.org/10.1145/2535838.2535874}}


\bibitem[\protect\citeauthoryear{F{\'e}votte and Lathuili{\`e}re}{F{\'e}votte
  and Lathuili{\`e}re}{2016}]%
        {verrou}
\bibfield{author}{\bibinfo{person}{Fran{\c c}ois F{\'e}votte} {and}
  \bibinfo{person}{Bruno Lathuili{\`e}re}.} \bibinfo{year}{2016}\natexlab{}.
\newblock \bibinfo{title}{{VERROU: Assessing Floating-Point Accuracy Without
  Recompiling}}.  (\bibinfo{date}{Oct.} \bibinfo{year}{2016}).
\newblock
\showURL{%
\url{https://hal.archives-ouvertes.fr/hal-01383417}}
\newblock
\shownote{working paper or preprint.}


\bibitem[\protect\citeauthoryear{Fousse, Hanrot, Lef\`evre, P\'elissier, and
  Zimmermann}{Fousse et~al\mbox{.}}{2007}]%
        {mpfr}
\bibfield{author}{\bibinfo{person}{Laurent Fousse}, \bibinfo{person}{Guillaume
  Hanrot}, \bibinfo{person}{Vincent Lef\`evre}, \bibinfo{person}{Patrick
  P\'elissier}, {and} \bibinfo{person}{Paul Zimmermann}.}
  \bibinfo{year}{2007}\natexlab{}.
\newblock \showarticletitle{{MPFR}: A Multiple-Precision Binary Floating-Point
  Library with Correct Rounding}.
\newblock \bibinfo{journal}{{\it {ACM} Trans. Math. Software}}
  \bibinfo{volume}{33}, \bibinfo{number}{2} (\bibinfo{date}{June}
  \bibinfo{year}{2007}), \bibinfo{pages}{13:1--13:15}.
\newblock
\showURL{%
\url{http://doi.acm.org/10.1145/1236463.1236468}}


\bibitem[\protect\citeauthoryear{Goubault and Putot}{Goubault and
  Putot}{2011}]%
        {fluctuat}
\bibfield{author}{\bibinfo{person}{Eric Goubault} {and} \bibinfo{person}{Sylvie
  Putot}.} \bibinfo{year}{2011}\natexlab{}.
\newblock \showarticletitle{Static Analysis of Finite Precision Computations}
  {\em (\bibinfo{series}{VMCAI'11})}. \bibinfo{publisher}{Springer-Verlag},
  \bibinfo{address}{Berlin, Heidelberg}, \bibinfo{pages}{232--247}.
\newblock
\showISBNx{978-3-642-18274-7}
\showURL{%
\url{http://dl.acm.org/citation.cfm?id=1946284.1946301}}


\bibitem[\protect\citeauthoryear{Higham}{Higham}{2002}]%
        {higham-book}
\bibfield{author}{\bibinfo{person}{Nicholas~J. Higham}.}
  \bibinfo{year}{2002}\natexlab{}.
\newblock \bibinfo{booktitle}{{\em Accuracy and Stability of Numerical
  Algorithms\/} (\bibinfo{edition}{2nd} ed.)}.
\newblock \bibinfo{publisher}{Society for Industrial and Applied Mathematics},
  \bibinfo{address}{Philadelphia, PA, USA}.
\newblock
\showISBNx{0898715210}


\bibitem[\protect\citeauthoryear{Jaeger}{Jaeger}{2016}]%
        {openlibm}
\bibfield{author}{\bibinfo{person}{Andreas Jaeger}.}
  \bibinfo{year}{2016}\natexlab{}.
\newblock \bibinfo{title}{OpenLibm}.
\newblock
\newblock
\showURL{%
\url{http://openlibm.org/}}


\bibitem[\protect\citeauthoryear{Kahan}{Kahan}{1965}]%
        {kahan-summation}
\bibfield{author}{\bibinfo{person}{W. Kahan}.} \bibinfo{year}{1965}\natexlab{}.
\newblock \showarticletitle{Pracniques: Further Remarks on Reducing Truncation
  Errors}.
\newblock \bibinfo{journal}{{\em Commun. ACM\/}} \bibinfo{volume}{8},
  \bibinfo{number}{1} (\bibinfo{date}{Jan.} \bibinfo{year}{1965}),
  \bibinfo{pages}{40--}.
\newblock
\showISSN{0001-0782}
\showDOI{%
\url{https://doi.org/10.1145/363707.363723}}


\bibitem[\protect\citeauthoryear{Kahan}{Kahan}{1971}]%
        {kahan-survey}
\bibfield{author}{\bibinfo{person}{William Kahan}.}
  \bibinfo{year}{1971}\natexlab{}.
\newblock \showarticletitle{A Survey of Error Analysis.}. In
  \bibinfo{booktitle}{{\em IFIP Congress (2)}}. \bibinfo{pages}{1214--1239}.
\newblock
\showURL{%
\url{http://dblp.uni-trier.de/db/conf/ifip/ifip71-2.html\#Kahan71}}


\bibitem[\protect\citeauthoryear{Kahan}{Kahan}{1987}]%
        {much-ado-nothing}
\bibfield{author}{\bibinfo{person}{W. Kahan}.} \bibinfo{year}{1987}\natexlab{}.
\newblock \showarticletitle{Branch Cuts for Complex Elementary Functions or
  Much Ado About Nothing's Sign Bit}.
\newblock In \bibinfo{booktitle}{{\em The State of the Art in Numerical
  Analysis (Birmingham, 1986)}}, \bibfield{editor}{\bibinfo{person}{A.~Iserles}
  {and} \bibinfo{person}{M.~J.~D. Powell}} (Eds.). \bibinfo{series}{Inst. Math.
  Appl. Conf. Ser. New Ser.}, Vol.~\bibinfo{volume}{9}.
  \bibinfo{publisher}{Oxford Univ. Press}, \bibinfo{address}{New York},
  \bibinfo{pages}{165–211}.
\newblock


\bibitem[\protect\citeauthoryear{Kahan}{Kahan}{1998}]%
        {no-sampling}
\bibfield{author}{\bibinfo{person}{W. Kahan}.} \bibinfo{year}{1998}\natexlab{}.
\newblock \bibinfo{booktitle}{{\em The Improbability of Probabilistic Error
  Analyses for Numerical Computations}}.
\newblock \bibinfo{type}{{T}echnical {R}eport}. \bibinfo{pages}{34} pages.
\newblock
\showURL{%
\url{http://www.cs.berkeley.edu/~wkahan/improber.pdf}}


\bibitem[\protect\citeauthoryear{Kahan}{Kahan}{2005}]%
        {kahan-future}
\bibfield{author}{\bibinfo{person}{William Kahan}.}
  \bibinfo{year}{2005}\natexlab{}.
\newblock \bibinfo{title}{Floating-Point Arithmetic Besieged by ``Business
  Decisions''}.
\newblock \bibinfo{howpublished}{World-Wide Web lecture notes.}. ,
  \bibinfo{numpages}{28}~pages.
\newblock
\showURL{%
\url{http://www.cs.berkeley.edu/~wkahan/ARITH_17.pdf}}


\bibitem[\protect\citeauthoryear{Lawson, Hanson, Kincaid, and Krogh}{Lawson
  et~al\mbox{.}}{1979}]%
        {blas}
\bibfield{author}{\bibinfo{person}{C.~L. Lawson}, \bibinfo{person}{R.~J.
  Hanson}, \bibinfo{person}{D.~R. Kincaid}, {and} \bibinfo{person}{F.~T.
  Krogh}.} \bibinfo{year}{1979}\natexlab{}.
\newblock \showarticletitle{Basic Linear Algebra Subprograms for Fortran
  Usage}.
\newblock \bibinfo{journal}{{\em ACM Trans. Math. Softw.\/}}
  \bibinfo{volume}{5}, \bibinfo{number}{3} (\bibinfo{date}{Sept.}
  \bibinfo{year}{1979}), \bibinfo{pages}{308--323}.
\newblock
\showISSN{0098-3500}
\showDOI{%
\url{https://doi.org/10.1145/355841.355847}}


\bibitem[\protect\citeauthoryear{Lee and Boehm}{Lee and Boehm}{1990}]%
        {boehm-fast}
\bibfield{author}{\bibinfo{person}{Vernon~A. Lee, Jr.} {and}
  \bibinfo{person}{Hans-J. Boehm}.} \bibinfo{year}{1990}\natexlab{}.
\newblock \showarticletitle{Optimizing Programs over the Constructive Reals}.
  In \bibinfo{booktitle}{{\em Proceedings of the ACM SIGPLAN 1990 Conference on
  Programming Language Design and Implementation}} {\em (\bibinfo{series}{PLDI
  '90})}. \bibinfo{publisher}{ACM}, \bibinfo{address}{New York, NY, USA},
  \bibinfo{pages}{102--111}.
\newblock
\showISBNx{0-89791-364-7}
\showDOI{%
\url{https://doi.org/10.1145/93542.93558}}


\bibitem[\protect\citeauthoryear{Martel}{Martel}{2009}]%
        {salsa-1}
\bibfield{author}{\bibinfo{person}{Matthieu Martel}.}
  \bibinfo{year}{2009}\natexlab{}.
\newblock \showarticletitle{Program Transformation for Numerical Precision}
  {\em (\bibinfo{series}{PEPM '09})}. \bibinfo{publisher}{ACM},
  \bibinfo{address}{New York, NY, USA}, \bibinfo{pages}{101--110}.
\newblock
\showISBNx{978-1-60558-327-3}
\showURL{%
\url{http://doi.acm.org/10.1145/1480945.1480960}}


\bibitem[\protect\citeauthoryear{McCullough and Vinod}{McCullough and
  Vinod}{1999}]%
        {distort-stock}
\bibfield{author}{\bibinfo{person}{B.~D. McCullough} {and}
  \bibinfo{person}{H.~D. Vinod}.} \bibinfo{year}{1999}\natexlab{}.
\newblock \showarticletitle{The Numerical Reliability of Econometric Software}.
\newblock \bibinfo{journal}{{\em Journal of Economic Literature\/}}
  \bibinfo{volume}{37}, \bibinfo{number}{2} (\bibinfo{year}{1999}),
  \bibinfo{pages}{633--665}.
\newblock


\bibitem[\protect\citeauthoryear{Minsky}{Minsky}{1967}]%
        {computable-reals}
\bibfield{author}{\bibinfo{person}{Marvin~L. Minsky}.}
  \bibinfo{year}{1967}\natexlab{}.
\newblock \bibinfo{booktitle}{{\em Computation: Finite and Infinite Machines}}.
\newblock \bibinfo{publisher}{Prentice-Hall, Inc.}, \bibinfo{address}{Upper
  Saddle River, NJ, USA}.
\newblock
\showISBNx{0-13-165563-9}


\bibitem[\protect\citeauthoryear{Nethercote and Seward}{Nethercote and
  Seward}{2007}]%
        {valgrind}
\bibfield{author}{\bibinfo{person}{Nethercote} {and} \bibinfo{person}{Seward}.}
  \bibinfo{year}{2007}\natexlab{}.
\newblock \showarticletitle{Valgrind: A Framework for Heavyweight Dynamic
  Binary Instrumentation}.
\newblock  (\bibinfo{date}{June} \bibinfo{year}{2007}).
\newblock


\bibitem[\protect\citeauthoryear{Ng}{Ng}{1993}]%
        {fdlibm}
\bibfield{author}{\bibinfo{person}{Dr. K-C Ng}.}
  \bibinfo{year}{1993}\natexlab{}.
\newblock \bibinfo{title}{FDLIBM}.
\newblock
\newblock
\showURL{%
\url{http://www.netlib.org/fdlibm/readme}}


\bibitem[\protect\citeauthoryear{Panchekha, Sanchez-Stern, Wilcox, and
  Tatlock}{Panchekha et~al\mbox{.}}{2015}]%
        {herbie}
\bibfield{author}{\bibinfo{person}{Pavel Panchekha}, \bibinfo{person}{Alex
  Sanchez-Stern}, \bibinfo{person}{James~R. Wilcox}, {and}
  \bibinfo{person}{Zachary Tatlock}.} \bibinfo{year}{2015}\natexlab{}.
\newblock \showarticletitle{Automatically Improving Accuracy for Floating Point
  Expressions}. In \bibinfo{booktitle}{{\em Proceedings of the 36th ACM SIGPLAN
  Conference on Programming Language Design and Implementation}} {\em
  (\bibinfo{series}{PLDI '15})}. \bibinfo{publisher}{ACM}.
\newblock


\bibitem[\protect\citeauthoryear{Plotkin}{Plotkin}{1970}]%
        {antiunification}
\bibfield{author}{\bibinfo{person}{Gordon~D. Plotkin}.}
  \bibinfo{year}{1970}\natexlab{}.
\newblock \showarticletitle{{A note on inductive generalization}}.
\newblock \bibinfo{journal}{{\em Machine Intelligence\/}}  \bibinfo{volume}{5}
  (\bibinfo{year}{1970}), \bibinfo{pages}{153--163}.
\newblock


\bibitem[\protect\citeauthoryear{Quinn}{Quinn}{1983}]%
        {wall-street-distort-stock}
\bibfield{author}{\bibinfo{person}{Kevin Quinn}.}
  \bibinfo{year}{1983}\natexlab{}.
\newblock \showarticletitle{Ever Had Problems Rounding Off Figures? {T}his
  Stock Exchange Has}.
\newblock \bibinfo{journal}{{\em The Wall Street Journal\/}}
  (\bibinfo{date}{November 8,} \bibinfo{year}{1983}), \bibinfo{pages}{37}.
\newblock


\bibitem[\protect\citeauthoryear{Shewchuk}{Shewchuk}{1996}]%
        {triangle}
\bibfield{author}{\bibinfo{person}{Jonathan~Richard Shewchuk}.}
  \bibinfo{year}{1996}\natexlab{}.
\newblock \showarticletitle{Triangle: {E}ngineering a {2D} {Q}uality {M}esh
  {G}enerator and {D}elaunay {T}riangulator}.
\newblock In \bibinfo{booktitle}{{\em Applied Computational Geometry: Towards
  Geometric Engineering}}, \bibfield{editor}{\bibinfo{person}{Ming~C. Lin}
  {and} \bibinfo{person}{Dinesh Manocha}} (Eds.). \bibinfo{series}{Lecture
  Notes in Computer Science}, Vol.~\bibinfo{volume}{1148}.
  \bibinfo{publisher}{Springer-Verlag}, \bibinfo{pages}{203--222}.
\newblock
\newblock
\shownote{From the First ACM Workshop on Applied Computational Geometry.}


\bibitem[\protect\citeauthoryear{Si}{Si}{2015}]%
        {tetgen}
\bibfield{author}{\bibinfo{person}{Hang Si}.} \bibinfo{year}{2015}\natexlab{}.
\newblock \showarticletitle{TetGen, a Delaunay-Based Quality Tetrahedral Mesh
  Generator}.
\newblock \bibinfo{journal}{{\em ACM Trans. Math. Softw.\/}}
  \bibinfo{volume}{41}, \bibinfo{number}{2}, Article \bibinfo{articleno}{11}
  (\bibinfo{date}{Feb.} \bibinfo{year}{2015}), \bibinfo{numpages}{36}~pages.
\newblock
\showISSN{0098-3500}
\showDOI{%
\url{https://doi.org/10.1145/2629697}}


\bibitem[\protect\citeauthoryear{Solovyev, Jacobsen, Rakamaric, and
  Gopalakrishnan}{Solovyev et~al\mbox{.}}{2015}]%
        {fptaylor}
\bibfield{author}{\bibinfo{person}{Alexey Solovyev}, \bibinfo{person}{Charlie
  Jacobsen}, \bibinfo{person}{Zvonimir Rakamaric}, {and}
  \bibinfo{person}{Ganesh Gopalakrishnan}.} \bibinfo{year}{2015}\natexlab{}.
\newblock \showarticletitle{Rigorous Estimation of Floating-Point Round-off
  Errors with Symbolic Taylor Expansions} {\em (\bibinfo{series}{FM'15})}.
  \bibinfo{publisher}{Springer}.
\newblock


\bibitem[\protect\citeauthoryear{Tange}{Tange}{2011}]%
        {gnu-parallel}
\bibfield{author}{\bibinfo{person}{O. Tange}.} \bibinfo{year}{2011}\natexlab{}.
\newblock \showarticletitle{GNU Parallel - The Command-Line Power Tool}.
\newblock \bibinfo{journal}{{\em ;login: The USENIX Magazine\/}}
  \bibinfo{volume}{36}, \bibinfo{number}{1} (\bibinfo{date}{Feb}
  \bibinfo{year}{2011}), \bibinfo{pages}{42--47}.
\newblock
\showURL{%
\url{http://www.gnu.org/s/parallel}}


\bibitem[\protect\citeauthoryear{{U.S. General Accounting Office}}{{U.S.
  General Accounting Office}}{1992}]%
        {patriot}
\bibfield{author}{\bibinfo{person}{{U.S. General Accounting Office}}.}
  \bibinfo{year}{1992}\natexlab{}.
\newblock \bibinfo{title}{Patriot Missile Defense: Software Problem Led to
  System Failure at Dhahran, Saudi Arabia}.
\newblock
\newblock
\showURL{%
\url{http://www.gao.gov/products/IMTEC-92-26}}


\bibitem[\protect\citeauthoryear{Wang, Zou, He, Xiong, Zhang, and Huang}{Wang
  et~al\mbox{.}}{2015}]%
        {fse16}
\bibfield{author}{\bibinfo{person}{Ran Wang}, \bibinfo{person}{Daming Zou},
  \bibinfo{person}{Xinrui He}, \bibinfo{person}{Yingfei Xiong},
  \bibinfo{person}{Lu Zhang}, {and} \bibinfo{person}{Gang Huang}.}
  \bibinfo{year}{2015}\natexlab{}.
\newblock \showarticletitle{Detecting and Fixing Precision-Specific Operations
  for Measuring Floating-Point Errors} {\em (\bibinfo{series}{FSE'15})}.
\newblock


\bibitem[\protect\citeauthoryear{Weber-Wulff}{Weber-Wulff}{1992}]%
        {round-elections}
\bibfield{author}{\bibinfo{person}{Debora Weber-Wulff}.}
  \bibinfo{year}{1992}\natexlab{}.
\newblock \bibinfo{title}{Rounding error changes Parliament makeup}.
\newblock
\newblock
\showURL{%
\url{http://catless.ncl.ac.uk/Risks/13.37.html\#subj4}}


\end{thebibliography}

\end{document}